%% file: main.tex
\documentclass[acmlarge]{acmart}

\AtBeginDocument{%
  \providecommand\BibTeX{{%
    \normalfont B\kern-0.5em{\scshape i\kern-0.25em b}\kern-0.8em\TeX}}}

\setcopyright{acmlicensed}
\acmJournal{TOSEM}
\acmYear{2023} \acmVolume{32} \acmNumber{1} \acmArticle{1} \acmMonth{1} \acmPrice{15.00}\acmDOI{10.1145/3583562}

\usepackage{algorithmic}
\usepackage{graphicx}
\usepackage{textcomp}
\usepackage{xcolor}
\def\BibTeX{{\rm B\kern-.05em{\sc i\kern-.025em b}\kern-.08em
    T\kern-.1667em\lower.7ex\hbox{E}\kern-.125emX}}
    
\usepackage{multirow}
\usepackage{hhline}
\usepackage{multicol}
\usepackage{lipsum}
\usepackage{colortbl}
\usepackage{lscape,float}

\usepackage{calc}
\usepackage{ifthen}
\usepackage{tikz}
\usepackage{pgf-pie}

\usepackage[lighttt]{lmodern}
\usepackage{listings}

\definecolor{codegreen}{rgb}{0,0.6,0}
\definecolor{codegray}{rgb}{0.5,0.5,0.5}
\definecolor{codepurple}{rgb}{0.58,0,0.82}
\definecolor{backcolour}{rgb}{0.95,0.95,0.92}
\definecolor{darkred}{rgb}{0.76, 0.13, 0.28}
 
\lstdefinestyle{mystyle}{
    backgroundcolor=\color{backcolour},   
    commentstyle=\color{codegreen},
    keywordstyle=\color{magenta},
    numberstyle=\tiny\color{codegray},
    stringstyle=\color{codepurple},
    basicstyle=\footnotesize,
    breakatwhitespace=false,         
    breaklines=true,                 
    captionpos=b,                    
    keepspaces=true,                 
    numbers=left,                    
    numbersep=5pt,                  
    showspaces=false,                
    showstringspaces=false,
    showtabs=false,                  
    tabsize=2
}
\newcommand{\toolname}{ToxiCR~}
\newcommand{\toolnameNS}{ToxiCR}

\newcommand{\code}[1]{{\tt #1}}

\newcommand{\revised}[1]{{#1}}
\newcommand{\excerpt}[1]{{\textcolor{purple}{\textit{#1}}}}

\newcommand{\significant}[1]{ \cellcolor{gray!25} #1}

\AtBeginDocument{%
  \providecommand\BibTeX{{%
    \normalfont B\kern-0.5em{\scshape i\kern-0.25em b}\kern-0.8em\TeX}}}
    
    \newenvironment{boxedtext}
    {
    
    \begin{center}

    \begin{tabular}{|p{0.96\linewidth}|}
    \hline
    }
    { 
    \\ \hline
    \end{tabular} 
    
    \end{center}
    \vspace{5pt}
       }

       \newcolumntype{L}[1]{>{\raggedright\let\newline\\\arraybackslash\hspace{0pt}}m{#1}}
\newcolumntype{C}[1]{>{\centering\let\newline\\\arraybackslash\hspace{0pt}}m{#1}}
\newcolumntype{R}[1]{>{\raggedleft\let\newline\\\arraybackslash\hspace{0pt}}m{#1}}
\usepackage{subfigure}

\begin{document}

\title{Automated Identification of Toxic Code Reviews Using ToxiCR}

\author{Jaydeb Sarker, Asif Kamal Turzo, Ming Dong, Amiangshu Bosu}

\email{jaydebsarker, asifkamal, mdong, amiangshu.bosu}
\affiliation{%
  \institution{Wayne State University}
  \city{Detroit}
  \state{Michigan}
  \country{USA}
}

\renewcommand{\shortauthors}{Sarker, et al.}

\begin{abstract}
 \input{Sections/abstract}

\end{abstract}


\maketitle

\input{Sections/introduction}
\input{Sections/background}
\input{Sections/research-prep}

\input{Sections/tool-design}

\input{Sections/evaluation}
\input{Sections/discussion}
\input{Sections/threats}
\input{Sections/conclusion}

\bibliographystyle{ACM-Reference-Format}
\bibliography{toxicity-references}

\end{document}

%% file: Sections/abstract.tex
Toxic conversations during software development interactions may have serious repercussions on a Free and Open Source Software (FOSS) development project. For example,  victims of toxic conversations may become afraid to express themselves, therefore get demotivated, and may  eventually leave the project. Automated filtering of toxic conversations may help a FOSS community to maintain healthy  interactions among its members.
However, off-the-shelf toxicity detectors perform poorly on Software Engineering (SE) dataset, such as one curated from code review comments. To encounter this challenge, we present \textit{ToxiCR}, a supervised learning-based toxicity identification tool for code review interactions. 
ToxiCR includes a choice to select one of the ten supervised learning algorithms, an option to select text vectorization techniques, eight {preprocessing} steps, and a large scale labeled dataset of 19,651 code review comments. {Two  out of those eight preprocessing steps are SE domain specific.}  With our rigorous evaluation of the models with various combinations of preprocessing steps and vectorization techniques, we have identified the best combination for our dataset that boosts 95.8\% accuracy and 88.9\% $F1_1$ score. ToxiCR significantly outperforms existing toxicity detectors on our dataset. We have released our dataset, pretrained models, evaluation results, and source code publicly available at: \url{https://github.com/WSU-SEAL/ToxiCR}).

%% file: Sections/introduction.tex
\section{Introduction}
\label{sec:intro} 

Communications among the members of many Free and Open Source Software(FOSS) communities include manifestations of toxic behaviours~\cite{squire2015floss,toxic-blog2,toxic-blog3,nafus2006gender,nasif-icse2019,paul-SANER-2019}. These toxic communications may have decreased the productivity of those communities by wasting valuable work hours~\cite{bosu2013impact,ramanstress}. FOSS developers being frustrated over peers with `prickly' personalities~\cite{Bosu-EMSE-2019-Blockchain,filippova2016effects}  may  contemplate leaving a community for good~\cite{de2004managing,toxic-blog}.  Moreover, as most FOSS communities rely on contributions from volunteers, attracting and retaining prospective joiners is crucial for the growth and survival of FOSS projects~\cite{qureshi2011socialization}. However, toxic interactions with existing members may pose barriers against successful onboarding of  newcomers~\cite{jensen2011joining,steinmacher2014support}. Therefore, it is crucial for FOSS communities to proactively identify and regulate toxic communications.

Large-scale FOSS communities, such as Mozilla, OpenStack, Debian, and GNU manage hundreds of projects and  generate large volumes of  text-based communications among their contributors. Therefore, it is highly time-consuming and infeasible for the project administrators to identify and timely intervene with ongoing toxic communications. Although, many FOSS communities have codes of conduct, those are rarely enforced due to time constraints ~\cite{toxic-blog}.  As a result, toxic interactions can be easily found within the communication archives of many well-known FOSS projects. As an anonymous FOSS developer wrote after leaving a toxic community, \revised{``}\textit{..it’s time to do a deep dive into the mailing list archives or chat logs. ... Searching for terms that degrade women (chick, babe, girl, bitch, cunt), homophobic slurs used as negative feedback (``that’s so gay”), and ableist terms (dumb, retarded, lame), may allow you to get a sense of how aware (or not aware) the community is about the impact of their language choice on minorities.}\revised{''}~\cite{toxic-blog}. Therefore, it is crucial to develop an automated tool to identify toxic communication of FOSS communities.

Toxic text classification is a Natural Language Processing (NLP) task to automatically classify a text as `toxic' or `non-toxic'.
There are several state-of-the-art tools to identify toxic contents in blogs and tweets~\cite{perspective-api, kurita2019towards, gunasekara2018review, alshemali2020improving}.  However, off-the-shelf toxicity detectors do not work well on Software Engineering (SE) communications~\cite{sarker2020apsec}, since several characteristics of such communications (e.g., code reviews and bug interactions) are different from those of blogs and tweets. For example, compared to code review comments,  tweets are shorter and are limited to a maximum length. Tweets rarely include SE domain specific technical jargon, URLs, or code snippets~\cite{ahmed2017senticr, sarker2020apsec}. Moreover, due to different meanings of some words (e.g, `kill', `dead', and `dumb') in the SE context, SE communications with such words are often incorrectly classified as `toxic' by off-the-shelf toxicity detectors~\cite{ramanstress,sarker2020apsec}.
 
 To encounter this challenge, Raman \textit{et} al. developed a toxicity detector tool (referred as the `STRUDEL
tool’ hereinafter) for the SE domain ~\cite{ramanstress}. However, as the STRUDEL tool was trained and evaluated with only 611 SE texts. {Recent studies have found that it performed poorly on new samples~\cite{qiu2022detecting,miller2022toxic}}. To further investigate these concerns, Sarker et al. conducted a benchmark study of the STRUDEL tool and four other off-the-shelf toxicity detectors using two large scale SE datasets{~\cite{sarker2020apsec}}. To develop their datasets, they empirically developed a rubric to determine which SE texts should be placed in the  `toxic' group during their manual labeling.  Using that rubric, they manually labeled a dataset of 6,533 code review comments and 4,140 Gitter messages~\cite{sarker2020apsec}. The results of their analyses suggest that none of the existing tools are reliable in identifying toxic texts from SE communications, since all the five
tools’ performances significantly degraded on their SE datasets. However, they also found noticeable performance boosts (i.e., accuracy improved from 83\% to 92\% and F-score improved from 40\% to 87\%) after retraining { two of the existing off-the-shelf models (i.e., DPCNN~\cite{zhang2015character} and BERT with FastAI~\cite{kurita2019towards})} using their datasets. Being motivated by these results, we hypothesize that a SE domain specific toxicity detector can boost even better performances, since off-the-shelf toxicity detectors do not use SE domain specific preprocessing steps, such as preprocessing of code snippets included within texts. On this hypothesis, this paper presents \toolnameNS, a SE domain specific toxicity detector. \toolname is trained and evaluated using a manually labeled dataset of 19,651 code review comments selected from four popular FOSS communities (i.e., Android, Chromium OS, OpenStack and LibreOffice).
{We selected code review comments, since a code review usually represents a direct interaction between two persons (i.e., the author and a reviewer). Therefore, a toxic code review comment has the potential to be taken as a personal attack and may hinder future collaboration between the participants.}
\toolname is written in Python using the Scikit-learn~\cite{scikit-learn} and TensorFlow~\cite{tensorflow}. It provides an option to train models using one of the ten supervised machine learning algorithms including five classical and ensemble-based, four deep neural network-based, and a Bidirectional Encoder Representations from Transformer (BERT) based ones. It also includes { eight preprocessing steps with two being SE domain specific and an option to choose from five different text vectorization techniques.}

We empirically evaluated various  optional preprocessing combinations for each of the ten algorithms to identify the best performing combination.
During our 10-fold cross-validations  evaluations, the best performing model of \toolname  significantly outperforms existing toxicity detectors on the code review dataset with an accuracy of 95.8\% and F-score of 88.9\%.


The primary contributions of this paper are:
\begin{itemize}
 \item \toolnameNS: An SE domain specific toxicity detector. \toolname is publicly available on Github at: \url{https://github.com/WSU-SEAL/ToxiCR}.  
 \item An empirical evaluation of ten machine learning algorithms to identify toxic SE communications.
 \item {Implementations of eight preprocessing steps including two SE domain specific ones that can be added to model training pipelines.}
 \item An empirical evaluation of three optional preprocessing steps in improving the performances of toxicity classification models.
 \item Empirical identification of the best possible combinations for all the ten algorithms.
\end{itemize}

\textbf{Paper organization:} The remainder of this paper is organized as following.
Section~\ref{background} provides a brief background and discusses prior related works. 
Section~\ref{sec:research-context} discusses the concepts utilized in designing ToxiCR.
Section~\ref{tool-design} details the design of \toolnameNS.
Section~\ref{sec:evaluation} details the results of our empirical evaluation.
Section~\ref{discussion} discusses the lessons learned based on this study. 
Section \ref{threats} discusses threats to validity of our findings. 
Finally, Section \ref{conclusion} provides a future direction based on this work and concludes this paper. 

%% file: Sections/background.tex
\section{Background}
\label{background}
This section defines toxic communications,  provides a brief overview of prior works on toxicity in FOSS communities and  describes state-of-the-art toxicity detectors.

\subsection{What constitutes a toxic communication?}
\label{sec:what-is-toxic}
Toxicity is a complex phenomenon to construct as it is highly subjective than other text classification problems (i.g., online abuse, spam) \cite{kumar2021designing}. Whether a communication should be considered as `toxic' also depends on a multitude of factors, such as  communication medium, location, culture, and relationship between the participants. In this research, we focus specially on written online communications. According to the Google Jigsaw AI team,  a text from an online communication can be marked as toxic if it contains disrespectful or rude comments that make a participant to leave the discussion forum \cite{perspective-api}. On the other hand, the Pew Research Center marks a text as toxic if it contains threat, offensive call, or sexually expletive words \cite{duggan2017online}. Anderson \textit{et} al.'s definition of toxic communication also includes insulting language or mockery \cite{anderson2018toxic}. Adinolf and Turkay studied toxic communication in online communities and their views of toxic communications include harassment, bullying, griefing (i.e, constantly making other players annoyed), and trolling~\cite{adinolf2018toxic}. 
To understand, how persons from various demographics perceive toxicity, Kumar \textit{et} al. conducted a survey with 17,280 participants inside the USA. To their surprise, their results indicate that  the notion toxicity cannot be attributed to any single demographic factor~\cite{kumar2021designing}. 
According to Miller et al., various antisocial behaviors fit inside the Toxicity umbrella such as hate speech, trolling, flaming, and cyberbullying \cite{miller2022toxic}. While some of the SE studies have investigated antisocial behaviors among SE communities using the `toxicity' construct~\cite{sarker2020apsec,ramanstress,miller2022toxic}, other studies have used various other lens such as incivility~\cite{ferreira2021shut}, pushback~\cite{egelman2020predicting}, and destructive criticism~\cite{gunawardena2022destructive}.  Table~\ref{tab:anti-social} provides a brief overview of the studied constructs and their definitions.
\begin{table}[]
    \centering
    \begin{tabular}{|p{2cm}|p{2cm}|p{10cm}|} \hline
    \textbf{ SE study} &    \textbf{Construct} &  \textbf{Definition} \\ \hline
    Sarker et al.~\cite{sarker2020apsec}      & Toxicity & ``\textit{includes any of the following: i) offensive name
calling, ii) insults, iii) threats, iv) personal attacks, v) flirtations, vi) reference to sexual activities, and vii) swearing or cursing.}'' \\  \hline
    Miller et al.~\cite{miller2022toxic}      & Toxicity & ``\textit{an umbrella term for various antisocial behaviors including trolling, flaming, hate speech, harassment,  arrogance, entitlement, and cyberbullying}''. \\ \hline
    
    Ferreira et al.~\cite{ferreira2021shut}      & Incivility & ``\textit{features of discussion that convey an unnecessarily disrespectful tone toward the discussion forum, its participants, or its topics}'' \\ \hline
    Gunawardena et al.~\cite{gunawardena2022destructive}      & Destructive criticism & \textit{negative feedback
which is nonspecific and is delivered in a harsh or sarcastic tone,  includes threats, or attributes poor task performance
to flaws of the individual.} \\ \hline
    Egelman et al.~\cite{egelman2020predicting} & Pushback & ``\textit{the perception of unnecessary interpersonal conflict in code review while a reviewer is blocking a change request}''\\ \hline
    \end{tabular}
    \caption{Anti-social constructs investigated in prior SE studies}
    \label{tab:anti-social}
\end{table}

\subsection{Toxic communications in FOSS communities}
 Several  prior studies have identified toxic communications in FOSS communities \cite{squire2015floss, ramanstress, sarker2020apsec, carillo2016towards,paul-SANER-2019}. Squire  and  Gazda  found occurrences of expletives and insults in publicly available IRC and mailing-list archives of top FOSS communities, such  as Apache,  Debian,  Django,  Fedora,  KDE,  and Joomla~\cite{squire2015floss}. More alarmingly, they identified sexist `maternal insults'  being used by many developers. 
 Recent studies have also reported toxic communications among issue discussions on Github~\cite{ramanstress}, and during code reviews~\cite{sarker2020apsec,paul-SANER-2019,gunawardena2022destructive,ferreira2021shut}. 
 
 Although toxic communications are rare in FOSS communities~\cite{sarker2020apsec}, toxic interactions can have severe consequences~\cite{carillo2016towards}. Carillo \textit{et} al. termed Toxic communications as a `poison' that impacts mental health of FOSS developers~\cite{carillo2016towards}, and may contribute to stress and burnouts~\cite{carillo2016towards,ramanstress}. When the level of toxicity increases in a FOSS community, the community may disintegrate as developers may no longer wish to be associated with that community~\cite{carillo2016towards}. Moreover, toxic communications hamper  onboarding of  prospective joiners, as a newcomer may get turned off by the signs of a  toxic   culture prevalent in a FOSS community~\cite{jensen2011joining,steinmacher2014support}. Miller \textit{et} al. conducted a qualitative study to better understand toxicity in the context of FOSS development~\cite{miller2022toxic}. They created a sample of 100 Github issues representing various  types of toxic interactions such as insults, arrogance, trolling, entitlement, and unprofessional behaviour. Their analyses also suggest that toxicity in FOSS communities differ from those observed on other online platforms such as Reddit or Wikipedia~\cite{miller2022toxic}.

 Ferreira et al. \cite{ferreira2021shut} investigated incivility during code  review discussions based on a qualitative analysis of 1,545 emails  from Linux Kernel Mailing Lists  and found that the most common forms of incivility among those forums are frustration, name-calling, and importance. 
Egelman et al. studied the negative experiences during code review, which they referred as ``pushback’’, a scenario when a reviewer is blocking a change request due to unnecessary conflict \cite{egelman2020predicting}. Qiu et al. further investigated  such ``pushback'' phenomena to automatically identify interpersonal conflicts~\cite{qiu2022detecting}. Gunawardena et al. investigated negative code review feedbacks based on a survey of 93 software developers, and they found that destructive criticism can be a threat to gender diversity in the software industry as women are less motivated to continue when they receive negative comments or destructive criticisms \cite{gunawardena2022destructive}.


\subsection{State of the art toxicity detectors}
To combat abusive online contents, Google's Jigsaw AI team developed the Perspective API (PPA), which is publicly available~\cite{perspective-api}. PPA is one of the general purpose state-of-the-art toxicity detectors. For a given text, PPA generates the probability (0 to 1) of that text  being toxic. 
As researchers are working to identify adversarial examples to deceive the PPA~\cite{hosseini2017deceiving}, the Jigsaw team  periodically updates it to eliminate  identified limitations. The Jigsaw team also published a guideline~\cite{jigsaw-guideline} to manually identify toxic contents and used that guideline to  curate a crowd-sourced labeled dataset of toxic online contents~\cite{kaggle2018}.
 This dataset has been used to train several deep neural network based toxicity detectors~\cite{georgakopoulos2018convolutional, chen2019use, elnaggar2018stop, srivastava2018identifying,gunasekara2018review,van2018challenges}. Recently, 
 Bhat \textit{et al.}  proposed \textit{ToxiScope},  a supervised learning-based classifier to identify toxic on workplace communications~\cite{bhat2021say}. However, ToxiScope's best model achieved a low F1-Score (i.e., =0.77) during their evaluation
 
 One of the major challenges in developing toxicity detectors is character-level obfuscations, where one or more characters of a toxic word are intentionally misplaced (e.g. fcuk),  or repeated (e.g., shiiit), or replaced (e.g., s*ck) to avoid detection.
To address this challenge,  researchers have used character-level encoders instead of word-level encoders to train neural networks~\cite{mishra2018neural, nobata2016abusive,kurita2019towards}. Although, character-level encoding based models can handle such character level obfuscations, they come with significant increments of computation times~\cite{kurita2019towards}.  
Several studies have also found racial and gender bias among contemporary toxicity detectors, as some trigger words (i.e., `gay', `black') are more likely to be associated with false positives (i.e, a nontoxic text marked as toxic)~\cite{vaidya2019empirical,xia2020demoting,sap2019risk}.

However, off-the-shelf toxicity detectors suffer significant performance degradation on SE datasets~\cite{sarker2020apsec}. Such degradation is not surprising, since prior studies found off-the-shelf natural language processing (NLP) tools also performing poorly on SE datasets~\cite{jongeling2017negative,ahmed2017senticr,novielli2018benchmark,lin2018sentiment}. Raman \textit{et} al. created the STRUDEL tool, an SE domain specific toxicity detector~\cite{ramanstress}, by leveraging the PPA tool and  a customized version of Stanford's Politeness Detector~\cite{danescu2013computational}. 
Sarker \textit{et} al. investigated the performance of the  STRUDEL tool and four other off-the-shelf toxicity detectors on two SE datasets  \cite{sarker2020apsec}. In their benchmark, none of the tools achieved reliable performance to justify  practical applications on SE datasets. However, they also achieved encouraging performance boosts, when they retrained two of the tools {(i.e., DPCNN~\cite{zhang2015character} and BERT with FastAI~\cite{kurita2019towards})} using their SE datasets.


%% file: Sections/research-prep.tex
\section{Research Context}
\label{sec:research-context}
To better understand our tool design, this section provides a brief overview of the machine learning (ML) algorithms integrated in \toolname and five word vectorization techniques for NLP tasks.

\subsection{Supervised machine learning algorithms }

For ToxiCR we selected ten supervised ML algorithms from the ones that have been commonly used for text classification tasks. Our selection includes three classical, two ensemble methods based, four deep neural network (DNN) based, and a Bidirectional Encoder Representations from Transformer (BERT) based algorithms. Following subsections provide a brief overview of the selected algorithms.

\subsubsection{Classical ML algorithms:}

We have selected  the following three classical algorithms, which have been previously used for classification of SE texts~\cite{ahmed2017senticr,calefato2017emotxt,kurtanovic2018user,terdchanakul2017bug,terdchanakul2017bug}.

\begin{enumerate}
    \item {Decision Tree (DT)} : 
    In this algorithm, the dataset is continuously split according to a certain parameter. DT has two entities, namely decision nodes and leaves. The leaves are the decisions or the final outcomes. And the decision nodes are where the data is split into two or more sub-nodes~\cite{quinlan1986induction}.
   
    \item { Logistic Regression (LR): } LR creates a mathematical model to predict the probability for one of the two possible outcomes and is commonly used for binary classification tasks~\cite{berkson1944application}. 
    
    \item { Support-Vector Machine (SVM):} After mapping the input vectors into a high dimensional non-linear feature space, SVM tries to identify the best hyperplane to partition the data into n-classes, where n is the number of possible outcomes
    ~\cite{cortes1995support}.  
\end{enumerate}

 \subsubsection{Ensemble methods:}
 Ensemble methods create multiple models and then combine them to produce improved results. We have selected  the following two ensemble methods based algorithms, based on prior SE studies~\cite{ahmed2017senticr,kurtanovic2018user,terdchanakul2017bug}.
 
 \begin{enumerate}
      \item {Random Forest (RF): } RF is an ensemble based method that  combines the results produced by multiple decision trees ~\cite{ho1995random}. RF creates independent decision trees and combines them in parallel using on the `bagging' approach~\cite{breiman1996bagging}. 
      
      \item {Gradient-Boosted Decision Trees (GBT):} Similar to RF, GBT is also an ensemble based method using decision trees \cite{friedman2001greedy}. However, GBT creates decision trees sequentially, so that each new tree can correct the errors of the previous one and combines the results using the `boosting' approach~\cite{schapire2003boosting}.
 \end{enumerate}

\subsubsection{Deep neural networks: }
In recent years, DNN based models have shown significant performance gains over both classical and ensemble based models in text classification tasks~\cite{zhang2015character,kowsari2017hdltex}. In this research, we have selected four  state-of-the-art DNN based algorithms.

\begin{enumerate}
    \item {Long Short Term Memory (LSTM)}:
    A Recurrent Neural Network (RNN) processes inputs sequentially,   remembers the past,  and  makes decisions based on what it has learnt from the past~\cite{rumelhart1986learning}. However, traditional RNNs may perform  poorly on long-sequence of inputs, such as those seen in text classification tasks due to `the vanishing gradient problem'. This problem occurs, when a RNN's weights are not updated effectively due to exponentially decreasing gradients. To overcome this limitation, Hochreiter and Schmidhuber proposed LSTM, a new type of RNN architecture, that overcomes the challenges posed by long term dependencies using a gradient-based learning algorithm~\cite{hochreiter1997long}. LSTM consists four units: i) input gate, which decides what information to add from current step, ii) forget gate, which decides what is to keep from prior steps,  iii) output gate, which determines the next hidden state, and iv) memory cell, stores information from previous steps.

\item {Bidirectional LSTM (BiLSTM):}
A BiLSTM is composed of a forward LSTM and a backward LSTM to model the input sequences more accurately than  an unidirectional LSTM \cite{graves2005framewise, cornegruta2016modelling}.
In this architecture, the forward LSTM takes input sequences in the forward direction to model information from the past, while the backward LSTM takes input sequences in the reverse direction to model information  from the future~\cite{huang2015bidirectional}.
BiLSTM has been shown to be performing better than the unidirectional LSTM in several text classification tasks, as it can identify language contexts better than LSTM~\cite{graves2005framewise}.

\item {Gated Recurrent Unit (GRU):}
Similar to LSTM, GRU belongs to the RNN family of algorithms. However, GRU aims to handle `the vanishing gradient problem' using a different approach than LSTM. GRU has a much simpler architecture with only two units, update gate and reset gate. 
The reset gate decides what information should be forgot for next pass and update gate determines which information should pass to next step. Unlike LSTM, GRU does not require any memory cell, and therefore needs shorter training time than LSTM~\cite{elsayed2019deep}.

\item {Deep Pyramid CNN (DPCNN):}
 Convolutional neural networks (CNN) are a specialized type of neural networks that utilizes a mathematical operation called convolution in at least one of their layers. CNNs are most commonly used for image classification tasks. 
Johnson and Zhang proposed a special type of CNN architecture, named deep pyramid convolutional neural network (DPCNN) for text classification tasks~\cite{johnson2017deep}. Although DPCNN achieves faster training time by utilizing word-level CNNs to represent input texts, it does not sacrifice accuracy over character-level CNNs due to its carefully designed  deep but low-complexity network architecture.

\end{enumerate}

\subsubsection{Transformer model:} In recent years,
Transformer based models have been used for sequence to sequence modeling such as neural machine translations. For a sequence to sequence modeling,  a Transformer architecture  includes two primary parts: i) \textit{the encoder}, which  takes the input and generates the higher dimensional vector representation, ii) \textit{the decoder}, which generates the the output sequence from the abstract vector from the encoder. For classification tasks, the output of encoders are used for training. Transformers solve the `vanishing gradient problem' on long text inputs using the `self attention mechanism', a technique to  identify the important features from different positions of an input sequence~\cite{vaswani2017attention}.

In this study, we select Bidirectional Encoder Representations from Transformers, which is commonly known as BERT~\cite{devlin2018bert}. 
BERT based models  have achieved remarkable performances in various NLP tasks, such as question answering, sentiment classification, and text summarization~\cite{adhikari2019docbert, devlin2018bert}.
BERT's transformer layers use multi-headed attention instead of recurrent units (e.g, LSTM, GRU) to model the contextualized representation of each word in an input.

\subsection{Word vectorization}
To train an NLP model, input texts need to be converted into a vector of features that machine learning models can work on. \textit{Bag-of-Words (BOW)} is one of the most basic representation techniques, that turns an arbitrary text into fixed-length vector by counting how many times each word appears. As BOW representations do not account for grammar and word order, ML models trained using BOW representations fail to identify relationships between words.
On the other hand,  \textit{word embedding} techniques convert words to n-dimensional vector forms in such a way that words having similar meanings have vectors close to each other in the n-dimensional space. Word embedding techniques can be further divided into two categories: i) context-free embedding, which  creates the same representation of a word  regardless of the context where it occurs; ii) contextualized word embeddings aim at capturing word semantics in different contexts to address the issue of polysemous (i.e., words with multiple meanings) and the context-dependent nature of words. 
For this research, we have experimented with five word vectorization techniques: one BOW based, three context-free, and one contextualized.  Following subsections provide a brief overview of those techniques.

\subsubsection{Tf-Idf:}
TF-IDF is a BOW based vectorization technique that evaluates how relevant a word is to a document in a collection of documents.  TF-IDF score for a word is computed by multiplying two metrics: how many times a word appears in a document ($Tf$), and the inverse document frequency of the word across a set of documents ($Idf$). Following equations show the computation steps for Tf-Idf scores.
\begin{equation}
    Tf\big(w,d\big)= f\big(w,d\big)/\sum\limits_{t\epsilon d}  f\big(t,d\big)
\end{equation}
Where,  $f\big(t,d\big)$ is the frequency of the word ($w$) in  the document ($d$), and  $\sum\limits_{t\epsilon d}  f\big(t,d\big)$ represents the total number of words in $d$. 
Inverse document frequency (Idf) measures the importance of a term across all documents.
\begin{equation}
    Idf \big(w\big)= log_e\big(N / w_N\big)
\end{equation}
Here, $N$  is the total number of documents and $w_N$ represents  the number of documents having w. Finally, we computed TfIdf score of a word as:
\begin{equation}
    TfIdf\big(w,d\big)= Tf\big(w,d\big) * Idf \big(w\big)
\end{equation}

\subsubsection{Word2vec:}  In 2013, Mikolaev \textit{et} al.~\cite{mikolov2013distributed} proposed Word2vec, a context free word embedding technique. It is based on two neural network models named Continuous Bag-of-Words (CBOW) and Skip-gram. CBOW predicts a target word based on its context, while skip-gram uses the current word to predict its surrounding context. During the training, word2vec takes a large corpus of text as input and generates a vector space, where \revised{each} word in the corpus is assigned a unique vector and words with similar meaning are located close to one another. 
        
\subsubsection{GloVe:}  
Proposed by Pennington et al.~\cite{pennington2014glove}\revised{,} Global Vectors for Word Representation (GloVe) is an unsupervised algorithm to create context free word embedding.
Unlike word2vec, GloVe generates vector space from global co-occurrence of words. 
      
\subsubsection{fastText:} Developed by  the Facebook AI team, \textit{fastText} is  a simple and efficient method to generate context-free word embeddings \cite{bojanowski2017enriching}. While Word2vec and GloVe cannot provide embedding for out of vocabulary words, fastText overcomes this limitation by taking into account morphological characteristics of individual words. A word's vector in fastText based embedding is built from vectors of substrings of characters contained in it. Therefore, fasttext performs better than Word2vec or GloVe in NLP tasks, if a corpus contains unknown or rare words~\cite{bojanowski2017enriching}.

 \subsubsection{BERT:} 
   
 Unlike context-free embeddings (e.g., word2vec, GloVe, and fastText), where  each word has a fixed representation regardless of the context within which the word appears,  a contextualized embedding  produces word representations that are dynamically informed by the words around them. In this study, we use BERT~\cite{devlin2018bert}.  Similar to fastText,  BERT can also handle out of vocabulary words.

%% file: Sections/tool-design.tex
\section{Tool Design} 
\label{tool-design}


    \begin{figure}[t]
	\centering  \includegraphics[width=14cm]{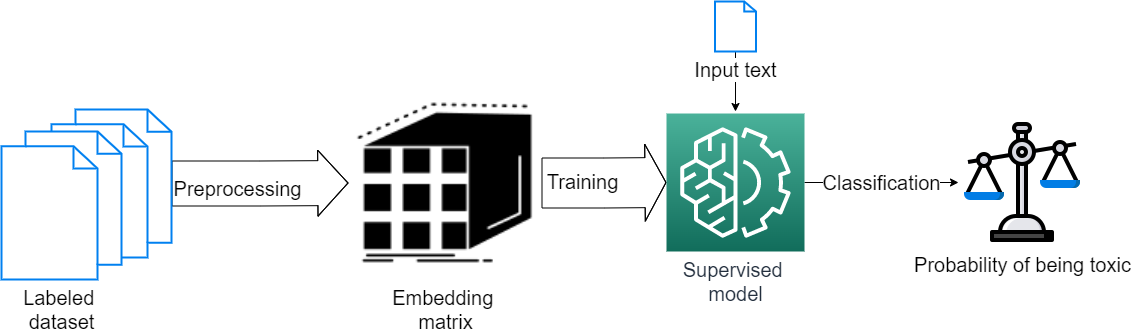}
	\caption{\small{ A simplified overview of ToxiCR showing key pipeline}}
	\label{fig:tool}	
	\vspace{-12pt}
\end{figure}

Figure~\ref{fig:tool} shows the architecture of ToxiCR. It takes a text ( i.e., code review comment) as input and applies a series of  mandatory preprocessing steps. Then, it applies a series of optional preprocessing based on selected configurations. Preprocessed texts are then fed into one of the selected vectorizers to extract features. Finally, output vectors are used to train and validate our supervised learning-based models.  The following subsections detail the research steps to design ToxiCR.

\subsection{Conceptualization of Toxicity}

{As we have mentioned in Section~\ref{sec:what-is-toxic}, what constitutes a `toxic communication' depends on various contextual factors. In this study, we specifically focus on popular FOSS projects such as Android, Chromium OS, LibreOffice, and OpenStack, where participants represent diverse culture, education, ethnicity, age, religion, gender, and political views. As participants are expected and even recommended to  maintain a high level of professionalism during their interactions with other members of those communities~\cite{openstack-coc,android-coc,libreoffice-coc}, we adopt the following expansive definition of toxic contents for this context.\footnote{ We introduced this definition in our prior study~\cite{sarker2020apsec}. We are repeating this definition to assist better comprehension of this paper's context}}:

\begin{quote}
 {\textcolor{darkred}{\textit{``{An SE conversation will be considered as toxic, if it includes any of the following: i) offensive name calling, ii) insults, iii) threats, iv) personal attacks, v) flirtations, vi) reference to sexual activities, and vii) swearing or cursing.''}}}}
\end{quote}

Our conceptualization of toxicity closely aligns with another recent work by Bhat \textit{et} el. that focuses on professional workplace communication \cite{bhat2021say}. According to their definition, a toxic behaviour includes any of the following:  sarcasm, stereotyping, rude statements, mocking conversations, profanity, bullying, harassment, discrimination and violence.

\subsection{Training Dataset Creation}
As of May 2021,  there are three  publicly available  labeled datasets of toxic communications from the SE domain. Raman et al.'s dataset created for the STRUDEL tool~\cite{ramanstress}  includes only 611 texts. In our recent benchmark study (referred as `the benchmark study' hereinafter), we created two datasets, i) a dataset of  6,533 code review comments selected from three popular FOSS projects (referred as `code review dataset 1' hereinafter),  i.e., Android, Chromium OS, and LibreOffice; ii) a dataset of 4,140 Gitter messages selected from the Gitter Ethereum channel (referred as `gitter  dataset' hereinafter)~\cite{sarker2020apsec}. We followed the exact same process used in the benchmark study to select and label additional 13,038 code review comments selected from the OpenStack projects.  In the following, we briefly describe our four-step process, which is detailed in our prior publication~\cite{sarker2020apsec}.

\subsubsection{Data Mining}
In the benchmark, we wrote a Python script  to mine the Gerrit~\cite{mukadam2013gerrit} managed code review repositories of three popular FOSS projects,  i.e., Android, Chromium OS, and LibreOffice.  Our script leverages Gerrit's REST API to mine and store all publicly available code reviews in a MySQL dataset. We use the same script to mine $\approx2.1$ million code review comments belonging to 670,996 code reviews from the OpenStack projects' code review repository hosted at \url{https://review.opendev.org/}.  We followed an approach similar to Paul \textit{et} al.~\cite{paul-SANER-2019} to identify potential bot accounts based on keywords (e.g., `bot', `auto', `build', `auto', `travis', `CI', `jenkins', and `clang'). If our manual manual validations of comments authored by a potential bot account confirmed it to be a bot, we excluded all comments posted by that account.

\subsubsection{Stratified sampling of code review comments}
 Since toxic communications are rare~\cite{sarker2020apsec} during code reviews, a randomly selected dataset of code review comments will be highly imbalanced with less than 1\% toxic instances. To overcome this challenge, we adopted a stratified sampling strategy as suggested by Särndal \textit{et} al. \cite{sarndal2003model}. We used Google's Perspective API (PPA)~\cite{perspective-api} to compute the toxicity score for each review comment. If the PPA score is more than 0.5, then the review comment is more likely to be toxic. Among the 2.1 million code review comments, we found 4,118 comments with PPA scores greater than 0.5. In addition to those 4,118 review comments, we selected 9,000 code review comments with PPA scores  less than 0.5. 
We selected code review comments with PPA scores less than 0.5 in a well distributed manner. We split the texts into five categories (i.e, score: 0-0.1, 0.11-0.2\revised{,} and so on) and took the same amount (1,800 texts) from each category. For example, we took 1,800 samples that have a score between 0.3 to 0.4.

 \begin{table}[]
\caption {An overview of the three SE domain specific toxicity datasets used in this study} \label{tab:dataset-overview} 
\centering
\begin{tabular}{|l|r|r|r|}
\hline
 \textbf{Dataset} & \textbf{\# total texts} &  \textbf{\# toxic} & \textbf{\# non-toxic} \\ \hline
   Code review 1  & 6,533 & 1,310 & 5,223 \\ \hline
  Code review 2  & 13,118 & 2,447 & 10,591 \\ \hline
  Gitter  dataset & 4,140 & 1,468 & 2,672        \\ \hline \hline
   Code review (combined)  & 19,651 & 3,757 & 15,819 \\ \hline

\end{tabular}
\end{table}

\subsubsection{Manual Labeling}


\begin{table*}
    \caption{{Rubric to label the SE text as toxic or non-toxic, adjusted from \cite{sarker2020apsec}}}
    \label{tab:rubrics}
    \centering
    \input{Tables/rubric}

\end{table*}

During the benchmark study \cite{sarker2020apsec},  we developed a manual labeling rubric fitting our definition as well as study context. Our initial rubric was based on the guidelines published by the Conversation AI Team (CAT)~\cite{jigsaw-guideline}. With these  guidelines as our starting point, two of the authors independently went through 1,000 texts to adopt the rules to better fit our context. Then, we had a discussion session to merge and create a unified set of rules. Table~\ref{tab:rubrics} represents our final rubric that has been used for manual labeling during both  the benchmark study and this study. 

Although we have used the guideline from the CAT as our starting point, our final rubric differs from the CAT rubric in two key aspects to better fit our target SE context. First, our rubric is targeted towards professional communities with contrast to the CAT rubric, which is targeted towards general online communications. Therefore, profanities and swearing to express a positive attitude may not be considered as toxic by the CAT rubric. For example, ``That's fucking amazing! thanks for sharing.'' is given as an example of ‘Not Toxic, or Hard to say’ by the CAT rubric. On the contrary,   any sentence with profanities or swearing  is considered `toxic' according to our rubric, since such a sentence does not constitute a healthy interaction.  Our characterization of profanities  also aligns with the recent SE studies on toxicity~\cite{miller2022toxic} and incivility~~\cite{ferreira2021shut}.  Second, the CAT rubric is for labeling on a four point-scale (i.e., `Very Toxic', `Toxic',  `Slightly Toxic or Hard to Say', and `Non toxic')~\cite{conversationai}. On the contrary, our labeling rubric is much simpler on a binary scale (`Toxic' and `Non-toxic'), since development of four point rubric  as well as classifier is significantly more challenging. We consider development of a four point rubric as a potential future direction.

Using this rubric, two of the authors independently labeled the 13,118 texts as either `toxic' or `non-toxic'. 
After the independent manual labeling, we compared the labels from the two raters to identify conflicts.  The two raters had agreements on 12,608 (96.1\%) texts during this process and achieved a Cohen's Kappa ($\kappa$) score of 0.92 (i.e., an almost perfect agreement)\footnote{Kappa ($\kappa$) values are commonly interpreted as follows:
values$\leq$ 0 as indicating ‘no agreement’ and 0.01 -- 0.20 as ‘none
to slight’, 0.21 -- 0.40 as ‘fair’, 0.41 -- 0.60 as ‘moderate’, 0.61--0.80
as ‘substantial’, and 0.81--1.00 as ‘almost perfect agreement’.}.  
We had meetings to discuss the conflicting labels and assign agreed upon labels for those cases. At the end of conflict resolution, we found 2,447 (18.76\%) texts labeled as `toxic' among the 13,118 texts. We refer to this dataset as `code review dataset 2' hereinafter. Table  \ref{tab:dataset-overview} provides an overview of the three dataset used in this study. 

\subsubsection{Dataset aggregation}
Since  the reliability of  a supervised learning-based model increases with the size of its training dataset, we decided to merge the two code review dataset into a single dataset (referred as `combined code review dataset' hereinafter). We believe such merging is not problematic, due to the following reasons. 
\begin{enumerate}
    \item Both of the datasets are labeled using the same rubrics and following the same protocol.
    \item We used the same set of raters for manual labeling.
    \item Both of the dataset are picked from the same type of repository (i.e., Gerrit based code reviews). 
\end{enumerate}

The merged code review dataset includes total 19,651 code review comments, where 3,757 comments (19.1\%) are labeled as `toxic'.

\subsection{Data preprocessing}
Code review comments are different from news, articles, books,
or even spoken language. For example, review comments often contain word contractions,  URLs, and code snippets. Therefore, we implemented eight data preprocessing steps. Five of those steps are mandatory, since those aim to remove unnecessary or redundant features. The remaining three steps are optional and their impacts on toxic code review detection are empirically evaluated in our experiments. {Two out of the three optional pre-processing steps are SE domain specific.} Table~\ref{tab:preprocessing} shows examples of  texts before and after preprocessing. 

\begin{table*}
    \caption{Examples of text preprocessing steps implemented in \toolname }
    \label{tab:preprocessing}
    \centering
    \input{Tables/preprocessing}
\end{table*}

\subsubsection{Mandatory preprocessing}
\toolname implements the following five mandatory pre-processing steps. 
\begin{itemize}

    \item \textit{URL removal (URL-rem):}  A code review comment may include an URL (e.g., reference to documentation or a StackOverflow post). Although URLs are irrelevant for a toxicity classifier, they can increase the number of features for supervised classifiers. We used a regular expression matcher to identify and remove all URLs from our datasets.
    
    \item \textit{Contraction expansion (Cntr-exp):} 
    Contractions, which are shortened form of one or two words, are common among code review texts. For example, some common words are: doesn't \textrightarrow does not, we're \textrightarrow we are.  By creating two different lexicons of the same term, contractions increase the number of unique lexicons and add redundant features. We replaced the commonly used 153 contractions, each with its expanded version.
    
    \item \textit{Symbol removal (Sym-rem):} Since special symbols (e.g., \&, \#, and \^{} ) are irrelevant for toxicity classification tasks, we use a regular expression matcher to identify and remove special symbols. 
    
    \item \textit{Repetition elimination (Rep-elm):} A person  may repeat some of the characters to misspell a toxic word to evade detection from a dictionary based toxicity detectors.  For example,    in the sentence ``You're duumbbbb!'', `dumb' is misspelled through character repetitions.   We have created a pattern based matcher to identify such misspelled cases and replace each with its correctly spelled form.
    
     \item \textit{Adversarial pattern identification (Adv-ptrn):} A person may misspell profane words by replacing some characters with a symbol (e.g., `f*ck' and `b!tch') or use an acronym for a slang (e.g., `stfu'). To identify such cases, we have developed a profanity preprocessor, which includes pattern matchers to identify various forms of the {85} commonly used profane words.  Our preprocessor replaces each identified case with its correctly spelled form. 
     
\end{itemize}

\subsubsection{Optional preprocessing}
\toolname includes options to apply following three  optional preprocessing steps. 

\begin{itemize}
\item \textit{Identifier splitting (Id-split):} In this preprocessing, we use a regular expression matcher to split identifiers written in both camelCase and under\_score forms. For example, this step will replace `isCrap' with `is Crap' and replace `is\_shitty' with `is shitty'.  This preprocessing may help to identify example code segments with profane words. 
    
\item \textit{Programming Keywords Removal (Kwrd-rem):}
Code review texts often include programming language specific keywords (e.g., `while', `case', `if', `catch', and `except'). These keywords are SE domain specific jargon and are not useful for toxicity prediction. We have created a list  of 90 programming keywords used in the popular programming languages (e.g., C++, Java, Python, C\#, PHP, JavaScript, and Go). This step searches and removes occurrences of those programming keywords from a text. 

\item \textit{Count profane words (profane-count):} Since the occurrence of profane words is suggestive of  a toxic text, we think  the number of profane words in a text may be an excellent feature for a toxicity classifier. We have created a list of 85 profane words, and this step counts the occurrences of these words in a text. While the remaining seven pre-processing steps modify an input text pre-vectorization,  this  step adds an additional dimension to the post vectored output of a text.

 \end{itemize}

\subsection{Word Vectorizers}
\toolname includes option to use five different word vectorizers. However, due to the limitations of the algorithms, each of the vectorizers can work with only one group of algorithms. In our implementation, Tfidf works only with the classical and ensemble (CLE) methods, Word2vec, GloVe, and fastText work with the deep neural network based algorithms, and BERT model includes its pre-trained vectorizer. For vectorizers, we chose the following implementations.

\begin{enumerate}
  
 \item  \textit{TfIdf:} We select the \code{TfidfVectorizer} from the scikit-learn library.  To prevent overfitting, we discard words not belonging to at least 20 documents in the corpus.
 
\item   \textit{Word2vec: } We select the pre-trained word2vec model available at: \url{https://code.google.com/archive/p/word2vec/}. This model was  trained with  a Google News dataset of 100 billion words and contains 300-dimensional vectors for 3 million words and phrases.

\item   \textit{GloVe: }
Among the publicly available, pretrained GloVe models (\url{https://github.com/stanfordnlp/GloVe}), we select the common crawl model.  This model was trained using web crawl data of 820 billion tokens and contains 300 dimensional vectors for 2.2 million words and phrases. 

\item   \textit{fastText:}
From the pretrained fastText models (\url{https://fasttext.cc/docs/en/english-vectors.html}), we select the common crawl model. This model was trained using the same dataset as our selected our GloVe model  and contains 300 dimensional vectors for 2 million words.

\item   \textit{BERT:}
We select a variant of BERT model published as `BERT\_en\_uncased'. 
This model was pre-trained on a dataset of 2.5 billion words from the Wikipedia and 800 million words from the Bookcorpus~\cite{Zhu_2015_ICCV}.

\end{enumerate}

\subsection{Architecture of the ML Models}
\label{sec-algos}
This section discusses the architecture of the ML models implemented in \toolnameNS. 

\subsubsection{Classical and ensemble methods}
We have used the scikit-learn~\cite{scikit-learn} implementations of the CLE classifiers. 

\begin{itemize}
    \item {Decision Tree (DT): } We have used  the \code{DecisionTreeClassifier}
    class with default parameters. 
    
    \item {Logistic Regression (LR): } We have used the \code{LogisticRegression} class with default parameters. 
    
    \item {Support-Vector Machine (SVM): } Among the various SVM implementations offered by scikit-learn, we have selected the \code{LinearSVC} class with default parameters. 
    
    \item {Random Forest (RF):} We have used the \code{RandomForestClassifier} class from scikit-learn ensembles. To prevent overfitting, we set the minimum number of samples to split at  to 5. For the other parameters, we accepted the default values. 
    
    \item {Gradient-Boosted Decision Trees (GBT):} We have used the \code{GradientBoostingClassifier} class from the scikit-learn library.  We set n\_iter\_no\_change =5, which stops the training early if  the last 5 iterations did not achieve  any improvement in accuracy. We accepted the default values for the other parameters.
\end{itemize}

\subsubsection{Deep Neural Networks Model}

We used the version 2.5.0 of the  TensorFlow library~\cite{tensorflow} for training the four deep neural network models (i.e., LSTM, BiLSTM, GRU, and DPCNN). 

Common parameters of the four models are:
\begin{itemize}
    \item We set \code{max\_features = 5000} (i.e., maximum number of features to use)  to reduce the memory overhead as well as to prevent model  overfitting.
    \item Maximum length of input is set to 500, which means our models can take texts with at most 500 words as inputs. Any input over this length would be truncated to 500 words.
    \item As all the three pre-trained word embedding models use  300 dimensional vectors to represent words and phrases, we have set embedding size to 300. 
    \item The embedding layer takes input embedding matrix as inputs.
    Each of word ($w_i$)  from a text is mapped (embedded) to a vector ($v_i$) using one of the three context-free vectorizers (i.e., fastText, GloVe, and word2vec). For a text $T$,  its embedding matrix  will have  a dimension of  $(300 X n)$,  where $n$ is the  total number of words in that text. 
   
    \item  Since we are developing binary classifiers, we have selected \code{binary\_crossentropy} loss function for model training.
    
    \item  We have selected the \code{Adam optimizer} (Adaptive Moment Estimation) \cite{kingma2014adam} to update the weights of the network during the training time. 
    The initial \code{learning\_rate} is set to  $0.001$. 
    
    \item During the training, we set \textit{accuracy} ($A$) as the evaluation metric. 
\end{itemize}

The  four deep neural models of ToxiCR are primarily based on three layers as  described briefly in the following. Architecture diagrams of the models are included in our replication package~\cite{replication}.

\begin{itemize}
    \item Input Embedding Layer: After preprocessing of code review texts, those are converted to input matrix. Embedded layer maps input matrix to a fixed dimension input embedding matrix. We used three pre-trained embeddings which help the model to capture the low level semantics using position based texts. 
    
    \item {Hidden State Layer}: This layer takes the position wise embedding matrix and helps to capture the high level semantics of words in code review texts. The configuration of this layer depends on the choice of the algorithm. ToxiCR includes one  CNN (i.e., DPCNN) and  three RNN (i.e., LSTM, BiLSTM, GRU) based hidden layers. In the following, we describe the key properties of these four types of layers.

    \begin{itemize}
    \item DPCNN blocks: Following the implementation of DPCNN \cite{johnson2017deep}, we set 7 convolution blocks with \code{Conv1D} layer after the input embedding layer.   We also set the other parameters of DPCNN model following ~\cite{johnson2017deep}. 
    Outputs from each of the CNN blocks is passed to a \code{GlobalMaxPooling1D} layer to capture the most important features from the inputs. A dense layer is set with 256 units which is activated with a linear activation function. 
    
        \item LSTM blocks:  From the Keras library, we use \textit{LSTM} unit to         capture the hidden sequence from input embedding vector. LSTM unit generates the high dimensional semantic representation vector. To reshape the output dimension, we use flatten and dense layer after LSTM unit.

        \item BiLSTM blocks: For text classification tasks, BiLSTM works better than LSTM for capturing the semantics of long sequence of text. Our model uses 50 units of Bidirectional LSTM units from the Keras library to generate the hidden sequence of input embedding matrix.  To downsample the high dimension hidden vector from BiLSTM units, we set a     \code{GlobalMaxPool1D} layer. This layer downsamples the hidden vector from BiLSTM layer by taking the maximum value of each dimension and thus captures the most important features for each vector.

        \item GRU blocks: We use bidirectional GRUs with 80 units to generate the hidden sequence of input embedding vector. To keep the most important features from GRU units, we set a concatenation of \code{GlobalAveragePooling1D} and \code{GlobalMaxPooling1D} layers. \code{GlobalAveragePooling1D} calculates the average of entire sequence of each vector and \code{GlobalMaxPooling1D} finds the maximum value of entire sequence. 
    \end{itemize}
    
    \item Classifier Layer: The output vector of hidden state layer project to the output layer with a dense layer and a \code{sigmoid} activation function. This layer generates the probability of the input vector from the range 0 to 1. We chose a \code{sigmoid} activation function because it provides the probability of a vector within 0 to 1 range.   
\end{itemize}

\subsubsection{Transformer models}

Among the several pre-trained BERT models\footnote{https://github.com/google-research/bert} we have used \code{bert\_en\_uncased}, which is also known as the $BERT\_base$ model. We downloaded the models from the \code{tensorflow\_hub}, which consists  trained machine learning models ready for fine tuning.

Our BERT model architecture is as following:
\begin{itemize}
    \item {Input layer:} takes the preprocessed input text from our SE dataset. To fit into BERT pretrained encoder, we preprocess each text using matching preprocessing model (i.e.  \code{bert\_en\_uncased\_preprocess}  \footnote{https://tfhub.dev/tensorflow/bert\_en\_uncased\_preprocess/3}).
    
    \item {BERT encoder:} From each preprocessed text, this layer produces BERT embedding vectors with higher level semantic representations. 
    
    \item {Dropout Layer:} To prevent overfitting as well as eliminate unnecessary features, outputs from the BERT encoder layer is passed to a dropout layer with a probability  of 0.1 to  drop an input. 
    
    \item {Classifier Layer:} Outputs from the  dropout layer is passed to a two-unit dense layer, which transforms the outputs into two-dimensional vectors. From these vectors, a one-unit dense layer with linear activation function generates the probabilities of each text being toxic. Unlike deep neural network's output layer, we have found that linear activation function provides better accuracy than non-linear ones (e.g, $relu$, $sigmoid$) for the BERT-based models.
    
    \item  {Parameters:} Similar to the deep neural network  models, {we} use \code{binary\_crossentropy} as the loss function and \textit{Binary Accuracy} as the evaluation metric during training.  
    
    \item {Optimizer:}  We set the optimizer as \code{Adamw} \cite{loshchilov2017decoupled} which improved the generalization performance of `adam' optimizer. \code{Adamw} minimizes the prediction loss and does regularization by decaying weight. Following the recommendation of Devlin \textit{et} al. ~\cite{devlin2018bert}, we set the initial learning rate to $3e-5$. 
    
\end{itemize}

\subsection{{ Model Training and Validation}}
Following subsections detail our model training and validation approaches.

\subsubsection{Classical and ensembles}
We evaluated all the models using 10-fold cross validations, where the dataset was randomly split into 10 groups and each of the ten groups was used as test dataset once, while the remaining nine groups were used to train the model. We used stratified split to ensure similar ratios of the classes  between the test and training sets.

\begin{table*}
    \caption{An overview of the hyper parameters for our deep neural networks and transformers }
    \label{tab:parameter}
    \centering
    \input{Tables/model_parameter}
\end{table*}

\subsubsection{DNN and Transformers}
We have customized several hyper-parameters of the DNN models to train our models. Table~\ref{tab:parameter} provides an overview of those customized hyper-parameters.
A DNN model can be overfitted due to over training. To encounter that, we have configured our training parameters to find the best fit model that is not overfitted.
During training, we split our dataset into three sets according to 8:1:1 ratio. These three sets are used for training, validation, and testing respectively during our 10-fold cross validations to evaluate our DNN and transformer models. For training, we have set maximum 40 epochs\footnote{the number times that a learning algorithm will work through the entire training dataset} for the DNN models and maximum 15 epochs for the BERT model. During each epoch, a model is trained using 80\% samples,  is validated using 10\% samples, and the remaining 10\% is used to measure the performance of the trained model. To prevent overfitting, we have used an \textit{EarlyStopping} function from the Keras library,  which monitors \textit{minimum val loss}. If the performance of a model on the validation dataset starts to degrade (e.g. loss begins to increase or accuracy begins to drop), then the training process is stopped.

\subsection{Tool interface}

We have designed ToxiCR to  support  standalone evaluation as well as being used as a library for toxic text identification. We have also included pre-trained models to save model training time. Listing~\ref{lst:usage} shows a sample code to  predict the toxicity of texts using our pretrained BERT model. 

  \begin{figure}[t!]
	\centering  \includegraphics[width=15cm]{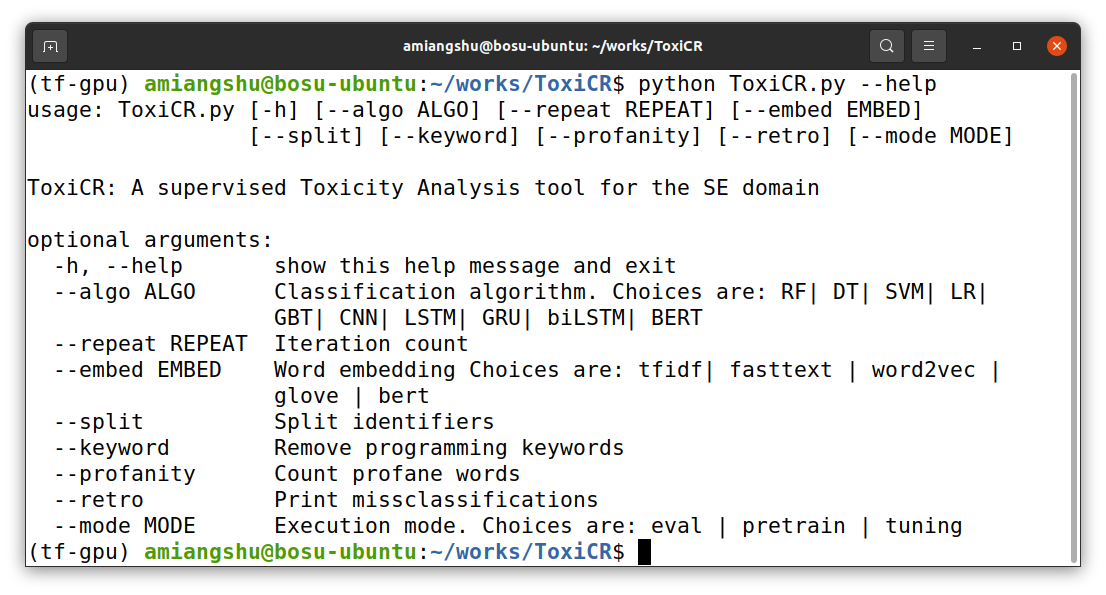}
	\caption{{The command line interface of ToxiCR showing various customization options}}
	\label{fig:tool_interface}	
	\vspace{-12pt}
\end{figure}

We have also included a command line based interface for model evaluation, retraining, and fine tuning hyperparameters.  Figure~\ref{fig:tool_interface} shows the usage help message of ToxiCR. Users can customize execution with eight optional parameters, which are as follows:

\begin{lstlisting} [language=Python, caption=Example usage of ToxiCR to classify toxic texts, label =lst:usage]
from ToxiCR import ToxiCR

clf=ToxiCR(ALGO="BERT", count_profanity=False, remove_keywords=True,
            split_identifier=False,
            embedding="bert", load_pretrained=True)

clf.init_predictor()
sentences=["this is crap", "thank you for the information",
            "shi*tty code" ]

results=clf.get_toxicity_class(sentences)

\end{lstlisting}

\begin{itemize}
    \item {Algorithm Selection:} Users can select one of the ten included algorithms by using the \code{--algo ALGO} option. 
    
    \item {Number of Repetitions:} Users can specify the number of times to repeat the 10-fold cross-validations in evaluation mode using \code{--repeat n} option. Default value is 5.
    
    \item {Embedding:} ToxiCR includes five different vectorization techniques: \code{tfidf}, \code{word2vec}, \code{glove}, \code{fasttext}, and \code{bert}. \code{tfidf} is configured to be used only with the CLE models. \code{word2vec}, \code{glove}, and \code{fastext} can be used only with the DNN models. Finally, \code{bert} can be used only with the transformer model. Users can customize this selection using the \code{--embed EMBED} option.
    
    \item {Identifier splitting:} Using the \code{--split} option, users can select to apply the optional preprocessing step to split identifiers written in camelCases or under\_scores.
    
    \item {Programming keywords:} Using the \code{--keyword} option, users can select to apply the optional preprocessing step to remove programming keywords.
    
    \item {Profanity:} The \code{--profanity} optional preprocessing step allows to add the number of profane words in a text as an additional feature.

    \item {Missclassification diagnosis:} The \code{--retro} option is useful for error diagnosis. If this option is selected, ToxiCR will write all misclassified texts in a spreadsheet to enable manual analyses.
    
    \item {Execution mode:} ToxiCR can be executed in three different modes. The \code{eval} mode will run 10-fold cross validations to evaluate the performance of an algorithm with the selected options. In the \code{eval} mode, ToxiCR  writes results of each run and model training time in a spreadsheet. The \code{retrain} mode will  train a classifier with the full dataset. This option is useful for saving models in a file to be used in the future. Finally, the \code{tuning} mode allows to explore various algorithm hyperparameters to identify the optimum set.
    
        \end{itemize}

%% file: Tables/rubric.tex
\resizebox{\textwidth}{!}{    
    \begin{tabular}{|l|p{4.4cm}|p{5.2cm}|p{3.3cm}|} \hline
     \textbf{\#} &\textbf{Rule}     & \textbf{Rationale} & \textbf{Example*} \\ \hline
    \textit{Rule 1:} & If a text includes  profane or curse words it would be marked as `toxic'.
     & Profanities are the most common sources of online toxicities.   & \excerpt{fuck! Consider it done!.} \\  \hline
  
    \textit{Rule 2:} & If a text includes an acronym, that generally refers to expletive or swearing, it would be marked as `toxic'.   & Sometimes people use acronyms of profanities, which are equally toxic as their expanded form.  & \excerpt{WTF are you doing!}  \\ \hline
   \textit{ Rule 3:} & Insulting remarks regarding another person or entities would be marked as `toxic'. & Insulting another developer may create a toxic environment and should not be encouraged.  & \excerpt{...shut up, smartypants.}\\ \hline
    
    \textit{Rule 4:} &Attacking  a person's identity  (e.g., race, religion, nationality, gender or sexual orientation) would be marked as `toxic'. & Identity attacks are considered toxic among all categories of online conversations. & \excerpt{Stupid fucking superstitious Christians.} \\ \hline
    
     \textit{ Rule 5:}&  Aggressive behavior or threatening another person or a community would be marked as `toxic'.  & Aggregations or threats may stir hostility between two developers and force the recipients to leave the community.   & \excerpt{yeah, but I'd really give a lot for an opportunity to punch them in the face.}  \\ \hline
    
    \textit{Rule 6:} & Both implicit or explicit References to  sexual activities would be marked as `toxic'. &  Implicit or explicit references to sexual activities may make some developers, particularly females, uncomfortable and make them leave a conversation. & \excerpt{This code makes me so horny. It's beautiful.} \\ \hline
    
     \textit{Rule 7:} & Flirtations would be marked as `toxic'. & Flirtations may also make a developer uncomfortable and make a recipient avoid the other person during future collaborations & \excerpt{I really miss you my girl}.   \\ \hline 
    
   \textit{ Rule 8:} & If a demeaning word (e.g., `dumb', `stupid', `idiot', `ignorant') refers to either the writer him/herself or his/her work, the sentence would not be marked as `toxic', if it does not fit any of the first seven rules.  & It is common in SE community to use those word for expressing their own mistakes. In those cases, the use of those toxic words to themselves or their does not make toxic meaning. While such texts are unprofessional~\cite{miller2022toxic}, those do not degrade future communication or collaboration.    &  \excerpt{I'm a fool and didn't get the point of the deincrement. It makes sense now.}  \\ \hline

\textit{ Rule 9:} & A sentence, that does not fit rules 1 through 8, would be marked as `non-toxic'. & General non-toxic comments.   & \excerpt{I think ResourceWithProps should be there instead of GenericResource.} \\  \hline

\multicolumn{4}{p{15cm}}{* \textit{Examples are provided verbatim from the datasets, to accurately represent the context. We did not censor any text, except omitting the reference to a  person's name.}}\\

    \end{tabular}
}

%% file: Tables/preprocessing.tex
\resizebox{\textwidth}{!}{
\begin{tabular}{|l|p{6cm}|p{6cm}|}
\hline
\textbf{Step}          & \textbf{Original}    & \textbf{Post Preprocessing} \\ \hline
URL-rem                    & ah crap. Not sure how I missed that. http://goo.gl/5NFKcD &  ah crap. Not sure how I missed that.                   \\ \hline

Cntr-exp          & this line shouldn't end with a period        & this line \textit{should not} end with a period            \\ \hline

Sym-rem       &  Missing: Partial-Bug: \#1541928                &   Missing  Partial Bug   1541928                \\ \hline

Rep-elm &  haha... \textit{looooooooser!}          &   haha.. loser!                 \\ \hline
Adv-ptrn & oh right, \textit{sh*t}   &     oh right, shit               \\ \hline

Kwrd-rem$\dagger$ & These \textit{static} values should be put at the top           &  These values should be put at the top                   \\ \hline

Id-split$\dagger$             & idp = self.\_create\_dummy\_idp (add\_clean\_up = False) & idp = self. create dummy idp(add clean up=  False) \\ \hline

\multicolumn{3}{l}{$\dagger$ -- \revised{an optional SE domain specific pre-processing step}}\\
\end{tabular}
}

%% file: Tables/model_parameter.tex
\begin{tabular}{|l|p{6cm}|l|}
\hline
\textbf{Hyper-Parameters} &\textbf{Deep neural networks (i.e., DPCNN, LSTM, BiLSTM, and GRU)} &  \textbf{Transformer (BERT)}               \\ \hline
Activation       & sigmoid  &   linear                \\ \hline
Loss function    & binary crossentropy  &  binary crossentropy  \\ \hline
Optimizer        & adam &     Adamw                 \\ \hline
Learning rate    & 0.001         & 3e-5              \\ \hline
Early stopping monitor & val\_loss &   val\_loss \\ \hline
Epochs & 40 &  15 \\ \hline
Batch size & 128 &   256 \\ \hline

\end{tabular}

%% file: Sections/evaluation.tex
\section{Evaluation}
\label{sec:evaluation}
We empirically evaluated the ten algorithms included in  \toolname to identify the best possible configuration to identify toxic texts from our datasets. Following subsections detail our experimental configurations and  the results of our evaluations. 

\subsection{Experimental Configuration}
To evaluate the performance of our models, we use precision, recall, f-score, accuracy for both toxic \revised{(class 1)} and non-toxic \revised{(class 0)} classes. 
We computed the following evaluation metrics.
 
 \begin{itemize}
     \item \textit{Precision ($P$): } For a class,  {precision}  is the percentage of identified cases that truly belongs to that class. 
     
     \item \textit{Recall ($R$):} For a class, {recall} is the ratio of correctly predicted cases and total number of  cases. 
     
     \item \textit{F1-score ($F1$):} F1-score  is the harmonic mean of precision and recall.  
     
     \item \textit{Accuracy ($A$):} {Accuracy}  is the percentage of cases that a model predicted correctly.
 \end{itemize}
 
In our evaluations, we consider F1-score for the toxic class (i.e., $F1_1$) as the most important metric to evaluate these models, since: i) identification of toxic texts  is our primary objective, and ii) our datasets are imbalanced with more than 80\% non-toxic texts.

To estimate the performance of the models more accurately, we repeated  10-fold cross validations five times and computed the means of all metrics over those 5 *10 =50 runs. We use Python's \code{Random} module, which is a pseudo-random number generator, to create stratified 10-fold partitions, preserving the ratio between the two classes across all partitions. If initialized with the same seed number, \code{Random} would generate the exact same sequence of pseudo-random numbers. At the start of each algorithm's evaluation, we initialized the \code{Random} generator using the same seed to ensure the exact same sequence of training/testing partitions for all algorithms. As the model performances are normally distributed, we use paired sample t-tests to check if observed performance differences between two algorithms are statistically significant ($p<0.05$). We use the `paired sample t-test', since our experimental setup guarantees   cross-validation runs of two different algorithms would get the same sequences of train/test partitions. We have included the results of the statistical tests in the replication package~\cite{replication}.

We conducted all evaluations on an Ubuntu 20.04 LTS workstation with Intel i7-9700 CPU, 32GB RAM, and an NVIDIA Titan RTX GPU with 24 GB memory. For python configuration, we created an Anaconda environment with Python 3.8.0, and \code{tensorflow}  / \code{tensorflow-gpu} 2.5.0.

\subsection{Baseline Algorithms}
To establish baseline performances, we computed the performances of four existing toxicity detectors (Table~\ref{tab:basline}) on our dataset. \revised{We briefly describe the four tools in the following.}  

\begin{enumerate}
    \item \revised{Perspective API~\cite{perspective-api} (off-the-shelf)}: 
    \revised{To prevent the online community from abusive content, Jigsaw and Google's Counter Abuse Technology team developed Perspective API~\cite{perspective-api}. Algorithms and datasets to train these models are not publicly available. Perspective API can generate the probability score of a text being toxic, servere\_toxic, insult, profanity, threat, identity\_attack, and sexually explicit. The score for each category is from 0 to 1 where the probability of a text belonging to that category increases with the score. For our two class classification, we considered a text as toxic if its Perspective API score for the toxicity category is higher than 0.5.} 
    
    \item \revised{STRUDEL tool~\cite{ramanstress} (off-the-shelf):} \revised{The STRUDEL tool is an ensemble based on  two existing tools: Perspective API and Stanford politeness detector and BoW vector obtained from preprocessed text. Its classification pipeline  obtains toxicity score of a text using the Perspective API,  computes politeness score using the Stanford politeness detector tool~\cite{danescu2013computational}, and computes BoW vector using TfIdf. For SE specificity,  its TfIdf vectorizer  excludes words that occur more frequently in the SE domain than in a non-SE domain. Although, STRUDEL tool also computes several other features such as sentiment score, subjectivity score, polarity score, number of LIWC anger words, and the number of emoticons in a text, none of these features contributed to improved performances during its evaluation~\cite{ramanstress}. Hence, the best performing ensemble from STRUDEL uses only the Perspective API score,  Stanford politeness score, and TfIdf vector. The off-the-shelf version is trained on a manually labeled dataset of 654 Github issues.}
    
    \item \revised{ STRUDEL tool~\cite{strudel-retrain} (retrain): 
    Due to several technical challenges, we were unable to retrain the STRUDEL tool using the source code provided in its repository~\cite{ramanstress}. Therefore, we wrote a simplified re-implementation based on the description included in the paper and our understanding of the current source code. Upon contacting, the primary author of the tool acknowledged our implementation as correct. Our implementation is publicly available inside the WSU-SEAL directory of the publicly available repository: \url{https://github.com/WSU-SEAL/toxicity-detector}. Our pull request with this implementation has been also merged to the original repository.  For computing baseline performance, we conducted a stratified 10-fold cross validation using our code review dataset.}

    \item \revised{DPCNN~\cite{sarker2020apsec} (retrain):} \revised{We cross-validated a DPCNN model  \cite{johnson2017deep}, using our code review dataset. We include this model in our baseline, since it provided the best retrained performance during our benchmark study~\cite{sarker2020apsec}.}
\end{enumerate}

\begin{table*}[t]
    \caption{Performances of the four contemporary toxic detectors to establish a baseline performance. For our classifications, we consider toxic texts as the `class 1' and non-toxic texts as the `class 0'. }
    \label{tab:basline}
    \centering
    \input{Tables/baseline_algorithm_performance}
\end{table*}

Table~\ref{tab:basline} shows the performances of the four baseline models. Unsurprisingly,  the two retrained models provide better performances than the off-the-shelf ones. Overall, the retrained Strudel tool provides the best scores among the four tools on all seven metrics. Therefore, we consider this model as the key baseline to improve on.
The best toxicity detector among the ones participating in the 2020 SemEval challenge achieved 0.92 $F_{1}$ score on the Jigsaw dataset~\cite{zaheri2020toxic}. As the baseline models listed in Table~\ref{tab:basline} are evaluated on a different dataset, it may not be fair to compare these models against the ones trained on Jigsaw dataset. However, the best baseline model's ${F_1}$ score is 7 (i.e., 0.92 -0.85 ) points lower than the ones from a non-SE domain. This result suggests  that with existing technology, it may be possible to train SE domain specific toxicity detectors with better performances than the best baseline listed in Table~\ref{tab:basline}.

\begin{boxedtext}
\textbf{{Finding 1:}} \emph{{Retrained models provide considerably better performances than the off-the-shelf ones, with the retrained STRUDEL tool providing the best performances.  Still the $F1_1$ score from the best baseline model lags 7 points behind the $F1_1$ score of state-of-the-art models trained and evaluated on the Jigsaw dataset during 2020 SemEval challenge~\cite{zaheri2020toxic}. }}
\end{boxedtext}

\subsection{How do the algorithms perform without optional preprocessing?}
Following subsections detail the performances of the three groups of algorithm described in the Section~\ref{sec-algos}.

\begin{table*}[!t]
    \caption{Mean performances of the ten selected algorithms based on  10-fold cross validations. For each group, shaded background indicate significant improvements over the others from the same group}
    \label{tab:acc2}
    \centering
    \input{Tables/accuracy_before_preprocessing}
\end{table*}

\subsubsection{Classical and Ensemble (CLE) algorithms}
The top five rows of the Table \ref{tab:acc2} (i.e., CLE group) show the performances of the five CLE models. Among, those five algorithms, RF achieves significantly higher $P_0$ (0.956), $F1_0$ (0.969), $R_1$ (0.81), $F1_1$ (0.859) and  accuracy (0.949) than the four other algorithms from this group. \revised{ The RF model also  significantly outperforms (One-sample t-test) the key baseline (i.e., retrained STRUDEL) in terms of the two key metrics, accuracy ($A$) and $F1_1$. Although, STRUDEL retrain achieves better recall ($R_1$), our RF based model achieves better precision ($P_1$).}

\subsubsection{Deep Neural Networks (DNN)}
\label{dnn-base}
We evaluated each of the four DNN algorithms  using three different pre-trained word embedding techniques (i.e., word2vec, GloVe, and fastText) to identify the best performing  embedding combinations. 
Rows 6 to 17 (i.e., groups: DNN1, DNN2, DNN3, and DNN4) of the Table \ref{tab:acc2} show the performances of the four DNN algorithms using three different embeddings.  For each group, statistically significant improvements (paired-sample t-tests) over the  other two configurations are highlighted using shaded background. 
Our results suggest that choice of embedding does influence performances of the DNN algorithms. However, such variations are minor. 

For DPCNN, only $R_1$ score is significantly better with fastText than it is with GloVe or word2vec. The other scores do not vary significantly among  the three embeddings. Based on these results, we recommend fastText for DPCNN in ToxiCR.
For LSTM and GRU, GloVe boosts significantly better $F1_1$ scores than those based on fastText or word2vec.   Since  $F1_1$ is one of the key measures to evaluate our models, we recommend the GloVe for both LSTM and GRU in ToxiCR. Glove also boosts the highest accuracy for both LSTM (although not statistically significant) and GRU. 
For BiLSTM, since fastText provides significantly higher $P_0$, $R_1$, and $F1_1$ scores than those based on GloVe or word2vec. We recommend fastText for BiLSTM in ToxiCR. These results also suggest that \revised{ three out of the four selected DNN algorithms (i.e., except LSTM) significantly outperform (one-sample t-test) the key baseline (i.e., retrained STRUDEL) in terms of both accuracy and $F1_1$-score.}

\subsubsection{Transformer}
The bottom row of Table~\ref{tab:acc2} shows the performance of our BERT based model. This model achieves the highest mean accuracy (0.957) and $F1_1$ (0.887) among all the 18 models listed in Table~\ref{tab:acc2}. \revised{This model also outperforms the baseline STRUDEL retrain on all the seven metrics.}

\begin{boxedtext}
\textbf{\revised{Finding 2:}} \emph{\revised{From the CLE group, RF provides the best performances. From the DNN group GRU with glove provides the best performances. Among the 18 models from the six groups, BERT achieves the best performances. Overall, ten out of the 18 models also outperform the baseline STRUDEL retrain model.}}
\end{boxedtext}

\subsection{ Do optional preprocessing steps improve performance?}
\label{sec:preprocess-eval}

For each of the ten selected algorithms, we evaluated whether the optional preprocessing steps (\revised{especially SE domain specific ones}) improve performances. Since ToxiCR includes  three optional preprocessing (i.e., identifier splitting (\textit{id-split}), keyword removal (\textit{kwrd-remove}), and counting profane words (\textit{profane-count}), we ran each algorithm with $2^3 = 8$ different combinations. For the DNN models, we did not evaluate all three embeddings in this step, as that would require evaluating \revised{3*8= 24} possible combinations for each one.   Rather we used only the best performing embedding identified in the previous step (i.e., Section~\ref{dnn-base}). 

To select the best optional preprocessing configuration from the eight possible configurations, we use mean accuracy and mean $F1_1$ scores based on five time 10-fold cross validations. We also used pair sampled t-tests to check whether any improvement over its base configuration's, as listed in the Table~\ref{tab:acc2} (i.e., no optional preprocessing selected), is statistical significant ( paired sample t-test, $p<0.05$).

Table~\ref{tab:acc3} shows the best  performing  configurations for all algorithms and the mean scores for those configurations.  Checkmarks ({\checkmark})  in the preprocessing columns for an algorithm indicate    that the best configuration for that algorithm does use that pre-processing. To save space, we report the performances of only the best combination for each algorithm. Detailed results are available in our replication package~\cite{replication}.

These results suggest that optional pre-processing steps do improve the performances of the models. Notably, CLE models gained higher improvements than the other two groups. RF's accuracy improved from 0.949 to 0.955 and $F1_1$ improved from 0.859 to   0.879 with the \textit{profane-count} preprocessing.

\begin{landscape}
    \vspace*{50pt}
\begin{table*}[htb!]
    \caption{Best performing configurations of each model with optional preprocessing steps. Shaded background indicates significant improvements over its base configuration (i.e., no optional preprocessing). For each column, bold font indicates the highest value for that measure. $\dagger$ -- \revised{ indicates an optional SE domain specific pre-processing step.}}
    \label{tab:acc3}
    \centering
    \input{Tables/accuracy_after_preprocessing}
\end{table*}
\end{landscape}

During these evaluations, other CLE models also achieved between 0.02 to 0.04 performance boosts in our key measures (i.e., $A$ and $F1_1$). Improvements from optional preprocessing also depend on algorithm choices. While the \textit{profane-count} preprocessing improved performances of all the CLE models, \textit{kwrd-remove} improved all except RF. On the other hand, \textit{id-split} improved none of the CLE models.

All the DNN models  also improved performances with the \textit{profane-count} preprocessing. Contrasting the CLE models, \textit{id-split}  was useful for three out of the four DNNs.  \textit{kwrd-remove} preprocessing improved only  LSTM models.
Noticeably, gains from optional preprocessing for the DNN models were less than 0.01 over the base configurations' and statistically insignificant (paired-sample t-test, $p>0.05$) for most of the cases.
Finally, although we noticed slight performance improvement (i.e., in $A$ and $F1_1$) of the BERT model with \textit{kwrd-remove}, the differences are not statistically significant. 
\textbf{Overall, at the end of our extensive evaluation, we found the best performing combination was a BERT model with \textit{kwrd-remove} optional preprocessing. The best combination provides 0.889 $F1_1$ score and 0.958 accuracy. } The best performing model also significantly  outperforms (one sample t-test, $p<0.05$)  the baseline model (i.e, \revised{STRUDEL retrain in Table~\ref{tab:basline}}) in all the seven performance measures.

\begin{boxedtext}
\textbf{\revised{Finding 3:}} \emph{\revised{Eight out of the ten models (i.e., except SVM and DPCNN) achieved significant performance gains through SE domain preprocessing such as programming keyword removal and identifier splitting. Although keyword removal may be useful for all the four classes of algorithms, identifier splitting is useful only for three DNN models. Our best model is based on BERT, which significantly outperforms the STRUDEL retrain model on all seven measures.}}
\end{boxedtext}

\subsection{How do the models perform on another dataset?}

To evaluate the generality of our models, we have used the Gitter dataset of 4,140 messages from our benchmark study~\cite{sarker2020apsec}. In this step, we conducted two types of evaluations. First, we ran 10-fold cross validations of the top CLE model (i.e., RF) and the BERT model using the Gitter dataset. Second, we evaluated cross dataset prediction performance \revised{(i.e., off-the-shelf)} by using the code review dataset for training and the Gitter dataset for testing. 

The top two rows of the Table~\ref{tab:cross_prediction_gitter} shows the results of 10-fold cross-validations for the the two models. We found that the BERT model provides the best accuracy (0.898) and the best $F1_1$ (0.856). On the Gitter dataset, all the seven performance measures achieved by the BERT model are lower than those on the  code review dataset. It may be due to the smaller size of the Gitter dataset (4,140 texts) than the code review dataset (19,651 texts). 
The bottom two rows of the Table~\ref{tab:cross_prediction_gitter} shows the results of our cross-predictions \revised{(i.e., off-the-shelf)}. Our BERT model achieved similar performances in terms of $A$ and $F1_1$ in both modes. However, the RF model performed better on the Gitter dataset in cross-prediction mode (i.e., off-the-shelf) than in cross-validation mode. This result further supports our hypothesis that the performance drops of our models on the Gitter dataset may be due to smaller sized training data.

\begin{table*}[!t]
    \caption{Performance of ToxiCR on Gitter dataset}
    \label{tab:cross_prediction_gitter}
    \centering
    \input{Tables/cross_prediction_gitter}
\end{table*}

\begin{boxedtext}
\textbf{\revised{Finding 4:}} \emph{\revised{Although, our best performing model provides higher precision off-the-shelf on  the Gitter dataset than that from the retrained model, the later achieves better recall. Regardless, our BERT model achieves similar accuracy and $F1_1$ during both off-the-shelf usage and retraining. }}
\end{boxedtext}

\subsection{What are the distributions of misclassifications from the best performing model?}

\begin{table*}[t]
    \caption{Confusion Matrix for our best performing model (i.e.,  BERT) for the  combined code review dataset}
    \label{tab:conf_matrix}
    \centering
    \input{Tables/confusion_matrix}
\end{table*}

The best-performing model (i.e., BERT) misclassified only 856 texts out of the 19,651 texts from our dataset. \revised{There are 373 false positives and 483 false negatives.} Table~\ref{tab:conf_matrix} shows the confusion matrix of the BERT model. To understand the reasons behind misclassifications, we adopted an open coding approach where two of the authors independently inspected each misclassified text to identify general scenarios. Next, they had a discussion session, where they developed an agreed upon higher level categorization scheme of five groups.  With this scheme, those two authors independently labeled each  misclassified text into one of those five groups. Finally, they compared their labels and resolved conflicts through mutual discussions.

    \begin{figure}[h!]
	\centering  \includegraphics[width=.9\textwidth]{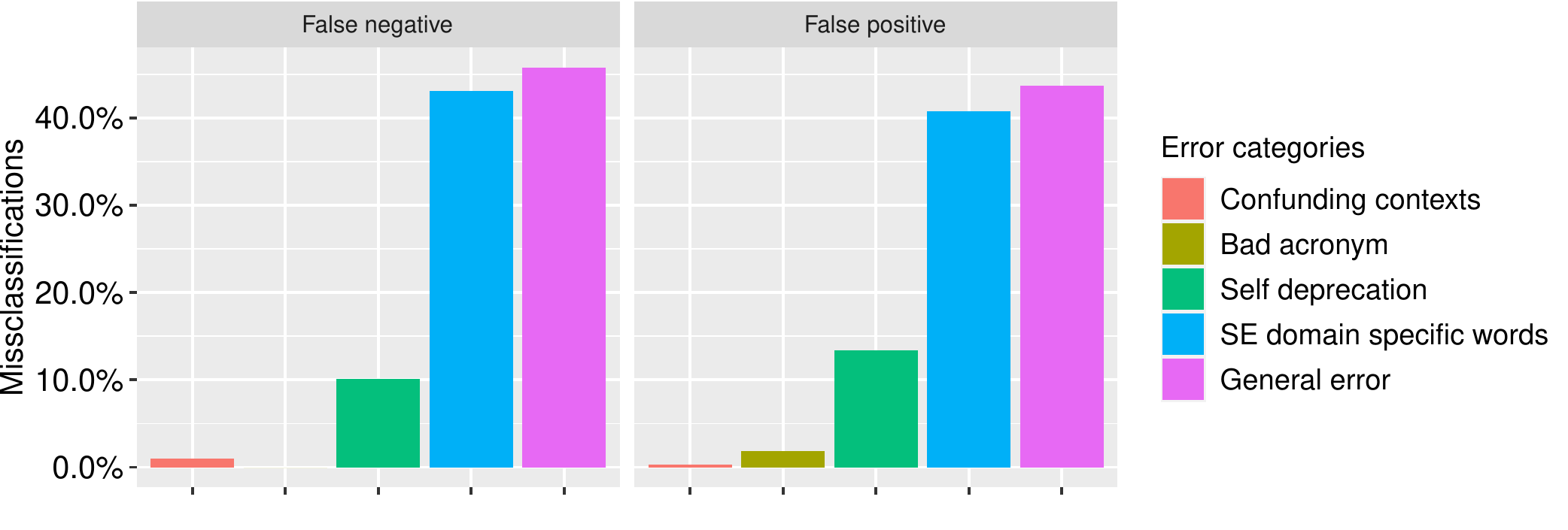}
	\caption{\small{Distribution of the misclassifications from the BERT model}}
	\label{fig:error_analysis}	
	\vspace{-12pt}
\end{figure}

Figure~\ref{fig:error_analysis} shows distributions of the five categories of misclassifcations from ToxiCR grouped by False Positives (FP) and False Negatives (FN). Following subsections detail those error categories.

\subsubsection{General erros (GE)}
General errors are due  to failures of  the classifier to identify the pragmatic meaning of various texts.
These errors represent 45\% of the false positives and 46\% of the false negatives. Many GE false positives are due to words or phrases that more frequently occur in toxic contexts and vice versa. 
For example, \excerpt{``If we do, should we just get rid of the HBoundType?''} and  \excerpt{``Done.  I think they came from a messed up rebase.''} are two false positive cases, due to the phrases `get rid of' and `messed up' that have occurred more frequently in toxic contexts. 

GE errors also occurred due to infrequent words.  For example,  \excerpt{``"Oh, look.  The stupidity that makes me rant so has already taken root. I suspect it's not too late to fix this, and fixing this rates as a mitzvah in my book."} -- is incorrectly predicted as non-toxic as a very few texts in our dataset include the word `stupidity'.  Another such instance was \excerpt{``this is another instance of uneducated programmers calling any kind of polymorphism overloading, please translate it to override.''}, due to to word `uneducated'.
As we did not have many instances of identify attacks in our dataset, most of those were also incorrectly classified. For example, \excerpt{``most australian dummy var name ever!"} was predicted as non-toxic by our classifier.

  \subsubsection{SE domain specific words (SE):}
     Words that have different meanings in the SE domain than its' meaning in the general domain (die, dead, kill, junk, and bug)~\cite{sarker2020apsec} were responsible for 40\% false positives and 43\% false negatives.  For example, the text \excerpt{``you probably wanted `die` here.  eerror is not fatal.''} , is incorrectly predicted as toxic due to the presence of the words `die' and `fatal'.  On the other hand,  although the word `junk' is used to  harshly criticize a code in the sentence \excerpt{``I don't actually need all this junk...''}, this sentence was predicted as non-toxic as most of the code review comments  from our dataset do not use `junk' in such a way.

     \subsubsection{Self deprecation (SD):}
     Usage of self-deprecating texts to  express humility is common during code reviews~\cite{sarker2020apsec,miller2022toxic}. \revised{We found that 13\% of 373 false positives and 11\% of 493 false negatives were due to the presence of self deprecating phrases.} For example, ``\excerpt{Missing entry in kerneldoc above... (stupid me)}'' is labeled as `non-toxic' in our dataset but is predicted as `toxic' by our model. Although, our model did classify many of the SD texts expressing humbleness correctly,  those texts also  led to some false negatives. For example, although   ``\excerpt{Huh? Am I stupid? How's that equivalent?}'' was misclassified as non-toxic, it fits `toxic' according to our rubric due to its aggressive tone.

    \subsubsection{Bad acronym (BA)} In few cases, developers have used acronyms with with alternate toxic expansion. For example,  the webkit framework used the acronym `WTF' -`Web Template Framework'\footnote{\url{https://stackoverflow.com/questions/834179/wtf-does-wtf-represent-in-the-webkit-code-base}}, for a namespace. Around 2\% of our false positive cases were  comments referring to the `WTF' namespace from Webkit.
    
    \subsubsection{Confounding contexts (CC)}
    Some of the texts in our dataset represent confounding contexts and were challenging even for the human raters to make a decision. Such cases represent 0.26\% false positives and 1.04\% false negatives. 
     For example, \excerpt{``This is a bit ugly, but this is what was asked so I added a null ptr check for |inspector\_agent\_|. Let me know what you think.''} is a false positive case from our dataset. We had labeled it as non-toxic, since the word `ugly' is applied to critique code written by the author of this text. 
    On the other hand, \excerpt{``I just know the network stack is full of \_bh poop. Do you ever get called from irq context? Sorry, I didn't mean to make you thrash.''} is labeled as toxic due to thrashing another person's code with the word `poop'. However, the reviewer also said sorry in the next sentence. During labeling, we considered it as toxic, since the reviewer could have critiqued the code in a nicer way. Probably due to the presence of mixed contexts, our classifier incorrectly predicted it as `non-toxic'.

\begin{boxedtext}
\textbf{{Finding 5:}} \emph{{Almost 85\% of the misclassifications are due to either our model's failure to accurately comprehend the pragmatic meaning of a text (i.e., GE) or words having SE domain specific synonyms.}}
\end{boxedtext}      

%% file: Tables/baseline_algorithm_performance.tex

\begin{tabular}{|l|r|r|r|r|r|r|r|}
\hline
 
\multirow{2}{*}{\textbf{Models}} & 

\multicolumn{3}{c|}{\textbf{Non-toxic}} & \multicolumn{3}{c|}{\textbf{Toxic}} & \multirow{2}{*}{\textbf{Accuracy}} \\ \cline{2-7}

  
  
 
 &   \textbf{$P_0$} & \textbf{$R_0$} & $F1_0$ & \textbf{$P_1$} & \textbf{$R_1$} & $F1_1$  & \\ \hline

  Perspective API~\cite{perspective-api} (off-the-shelf) &  0.92 & 0.79 & 0.85  & 0.45  & 0.70 & 0.55 & 0.78     \\ \hline
 Strudel Tool (off-the-shelf)~\cite{ramanstress} &   0.93 & 0.76 & 0.83 & 0.43 & {0.77} & 0.55  & 0.76   \\ \hline
 
 \revised{Strudel (retrain)~\cite{strudel-retrain}} & \revised{\textbf{0.97}} & \revised{\textbf{0.96}} & \revised{\textbf{0.97}} & \revised{\textbf{0.85}} & \revised{\textbf{0.86}} & \revised{\textbf{0.85}}  & \revised{\textbf{0.94}}   \\ \hline
  DPCNN (retrain) \cite{sarker2020apsec}  & {0.94}  & 0.95  & {0.94 }  &{ 0.81} & 0.76  &  {0.78}  & {0.91}  \\ \hline

\end{tabular}


%% file: Tables/accuracy_before_preprocessing.tex
\resizebox{\textwidth}{!}{    

\begin{tabular}{|l|l|c|r|r|r|r|r|r|r|}
\hline
\multirow{2}{*}{\textbf{Group}} &
\multirow{2}{*}{\textbf{Models}} & \multirow{2}{*}{\textbf{ Vectorizer}} & \multicolumn{3}{c|}{\textbf{Non-toxic}} & \multicolumn{3}{c|}{\textbf{Toxic}} & \multirow{2}{*}{\textbf{Accuracy ($A$)}} \\ \cline{4-9}

  
  
 
& & &  \textbf{$P_0$} & \textbf{$R_0$} & $F1_0$ & \textbf{$P_1$} & \textbf{$R_1$} & $F1_1$  & \\ \hline

  \multirow{5}{*}{{CLE}}  & DT & tfidf   & 0.954 & 0.963 & 0.959 & 0.841 & 0.806 & 0.823 & 0.933  \\ \cline{2-10}
  
  & GBT  & tfidf & 0.926 & \significant{0.985} & 0.955 & 0.916 & 0.672 & 0.775 & 0.925  \\ \cline{2-10}

  & LR  & tfidf & 0.918 & {0.983} & 0.949 & 0.900 & 0.633 & 0.743 & 0.916  \\ \cline{2-10}
   
    & RF & tfidf &  \significant{0.956} & 0.982 & \significant{0.969} & {0.916} & \significant{0.810} & \significant{0.859} & \significant{0.949}    \\ \cline{2-10}
 
  &SVM  & tfidf & 0.929 & 0.979 & 0.954 & 0.888 & 0.688 & 0.775 & 0.923  \\ 
  
 \hline \hline

  \multirow{3}{*}{{DNN1}}&  DPCNN & word2vec & 0.962 & 0.966  & 0.964 & 0.870 & 0.841 & 0.849 & 0.942    \\ \cline{2-10}
  
  &DPCNN & GloVe &  0.963 & 0.966 & 0.964 & {0.871} & 0.842 & 0.851 & 0.943  \\ \cline{2-10}

  & DPCNN & fasttext & {0.964} & {0.967} & {0.965} & 0.870 & \significant{0.845} & {0.852} & {0.944}  \\ \hline \hline
  
   \multirow{3}{*}{{DNN2}}& LSTM  & word2vec & 0.929 & \significant{0.978} & 0.953 & {0.866} & 0.698 & 0.778 & 0.922  \\ \cline{2-10}
  
  &LSTM  & GloVe & \significant{0.944} & 0.971 & {0.957} & 0.864 & \significant{0.758} & \significant{0.806} & 0.930  \\ \cline{2-10}
  
  &LSTM  & fasttext & 0.936  & 0.974  & 0.954 & 0.853 & 0.718 & 0.778 & 0.925  \\ \hline  \hline

   \multirow{3}{*}{{DNN3}}& BiLSTM & word2vec & 0.965 & 0.974 & {0.969} & 0.887 & 0.851 & 0.868 & 0.950  \\ \cline{2-10}
  
  &BiLSTM & GloVe & 0.965 & {0.976} & 0.968 & {0.895} & 0.828 & 0.859 & 0.948  \\ \cline{2-10}
  
  &BiLSTM & fasttext & \significant{0.966} & 0.974 & 0.970 & 0.888 & \significant{0.854} & \significant{0.871} & {0.951}  \\ \hline \hline
  
   \multirow{3}{*}{{DNN4}}& GRU & word2vec  & 0.964 & 0.976 & 0.970 & 0.894 & 0.847 & 0.870 & 0.951  \\ \cline{2-10}
  
  &GRU & GloVe  & 0.965 & \significant{0.977} & \significant{0.971} & \significant{0.901} & 0.851 & \significant{0.875} & \significant{0.953}  \\ \cline{2-10}

   &GRU & fasttext  & {0.965} & 0.974 & 0.969 & 0.888 & {0.852} & 0.869 & 0.951  \\ \hline \hline
   
    Transformer&  BERT & en  uncased & 0.971 & 0.976 & 0.973 & 0.901 & 0.876 & {0.887} & 0.957  \\ \hline

\end{tabular}

}

%% file: Tables/accuracy_after_preprocessing.tex

\begin{tabular}{|l|c|c|C{1.3cm}|C{1.2cm}|C{1cm}|r|r|r|r|r|r|R{1cm}|}
\hline
 \multirow{2}{*}{{Group}}  & \multirow{2}{*}{\textbf{Algo}} & \multirow{2}{*}{\textbf{Vectorizer}} &  \multicolumn{3}{|c|}{ \textbf{Preprocessing}} & \multicolumn{3}{|c|}{\textbf{Non-toxic}} &  \multicolumn{3}{|c|}{\textbf{Toxic}} & \multirow{2}{*}{\textbf{$A$}} \\
 \hhline{~~~---------~}
& & & \textbf{profane-count} & \textbf{kwrd-remove} & \textbf{id-split} &    \textbf{$P_0$} & \textbf{$R_0$} & $F1_0$ & \textbf{$P_1$} & \textbf{$R_1$} & $F1_1$  & \\  \hline
 
\multirow{5}{*}{{CLE}}  &   DT &  tfidf & \checkmark & \checkmark & - & \significant{0.960} & \significant{0.96}8 & \significant{0.964} & \significant{0.862} & \significant{0.830} & \significant{0.845} & \significant{0.942}  \\ \hhline{~~-----------}

  &GBT & tfidf  &    \checkmark &  \checkmark & - & \significant{0.938} & \significant{0.981} & \significant{0.959} & \significant{0.901} & \significant{0.729} & \significant{0.806} & \significant{0.932}  \\ \hhline{~-----------}

  &LR & tfidf   & \checkmark  & \checkmark & - & \significant{0.932} & \significant{0.981} & \significant{0.956} & 0.898 & \significant{0.698} & \significant{0.785} & \significant{0.927}  \\  \hhline{~-----------}
  
  & RF    &  tfidf & \checkmark & - & - & \significant{0.964} & \textbf{0.981} & \significant{0.972} & \textbf{0.917} &\significant{ 0.845} & \significant{0.879} &  \significant{0.955}    \\ \hhline{~-----------}
  
  &SVM &  tfidf &  \checkmark & \checkmark & - & \significant{0.939} & \significant{0.977} & \significant{0.958} & 0.886 & \significant{0.736} & \significant{0.804} & \significant{0.931}  \\ \hline \hline
  
  \multirow{4}{*}{{DNN}}  &DPCNN &fasttext &   \checkmark & - & - & 0.964 & 0.973 & \significant{0.968} & 0.889 & 0.846 & 0.863 & \significant{0.948}  \\ \hhline{~-----------}
  
  &LSTM & glve&   \checkmark &  \checkmark &  \checkmark & 0.944   & \significant{0.974} & \significant{0.959} & \significant{0.878} & 0.756 & 0.810 & \significant{0.932}  \\ \hhline{~-----------}   
  
  &BiLSTM &  fasttext &  \checkmark & - & \checkmark & 0.966 & 0.975 & \significant{0.971} & 0.892 & 0.858 & \significant{0.875} & \significant{0.953}  \\ \hhline{~-----------}
  
   &BiGRU & glove&   \checkmark & - & \checkmark   & 0.966 & 0.976 & 0.971 & 0.897 & \significant{0.856} & {0.876} & 0.954  \\ \hline \hline
   
   Transormer &BERT &bert &   - & \checkmark & -   & \textbf{0.970} & 0.978 & \textbf{0.974} & 0.907 & \textbf{0.874} & {\textbf{0.889}} & \textbf{0.958}  \\ \hline

\end{tabular}


%% file: Tables/cross_prediction_gitter.tex
\resizebox{\textwidth}{!}{    

\begin{tabular}{|p{2.5cm}|l|c|r|r|r|r|r|r|r|}
\hline
 
\multirow{2}{*}{\textbf{Mode}} &\multirow{2}{*}{\textbf{Models}} & \multirow{2}{*}{\textbf{ Vectorizer}} & \multicolumn{3}{l|}{\textbf{Non-toxic}} & \multicolumn{3}{l|}{\textbf{Toxic}} & \multirow{2}{*}{\textbf{Accuracy}} \\ \hhline{~~~------~}

  
  
 
 & & &  \textbf{$P$} & \textbf{$R$} & $F1$ & \textbf{$P$} & \textbf{$R$} & $F1$  & \\ \hline
 
\multirow{2}{3.4cm}{{Cross-validation \revised{(retrain)}}} &RF & TfIdf & 0.851    & 0.945   &    0.897 & 0.879  & 0.699   & 0.779   & 0.859      \\ \hhline{~---------}
 
  & BERT & BERT-en-uncased & 0.931    & 0.909     & 0.919  & 0.843   & 0.877  & 0.856  & 0.898     \\ \hline \hline
  \multirow{2}{3.4cm}{{Cross-prediction \revised{(off-the-shelf)}}} & RF & TfIdf & 0.857    & 0.977   &    0.914 & 0.945  & 0.704   & 0.807   & 0.881      \\ \hhline{~---------}
 
 
  &BERT & BERT-en-uncased & 0.897    & 0.949    & 0.923  & 0.897   & 0.802  & 0.847  & 0.897\\ \hline

\end{tabular}

}

%% file: Tables/confusion_matrix.tex
\begin{tabular}{l|l|c|c|c}
\multicolumn{2}{c}{}&\multicolumn{2}{c}{Predicted}&\\
\cline{3-4}
\multicolumn{2}{c|}{}&Toxic&Non-toxic&\multicolumn{1}{c}{}\\
\cline{2-4}
\multirow{2}{*}{Actual}& Toxic & $3259 $ & $483$ & \\
\cline{2-4}
& Non-toxic & $373$ & $15,446$ & \\
\cline{2-4}

\end{tabular}

%% file: Sections/discussion.tex
\section{Implications} \label{discussion}

Based on our design and evaluation of ToxiCR, we have identified following lessons.

\vspace{4pt} \noindent \textbf{Lesson 1: Development of a reliable toxicity detector for the SE domain is feasible.}
Despite of creating an ensemble of multiple NLP models (i.e., Perspective API, Sentiment score, Politeness score, Subjectivity, and Polarity) and various categories of features (i.e., BoW, number of anger words, and emoticons), the STRUDEL tool achieved only 0.57 F-score during their evaluation. Moreover, a recent study by Miller \textit{et} al. found false positive rates as high as 98\%~\cite{miller2022toxic}. On the other hand, the best model from the `2020 Semeval Multilingual Offensive Language Identification
in Social Media task' achieved a $F1_1$ score of 92.04\%~\cite{zampieri2020semeval}. Therefore, the question remains, whether we can build a SE domain specific toxicity detector that achieves similar performances (i.e., $F_1$ =0.92) as the ones from non-SE domains.

In designing ToxiCR, we adopted a different approach, i.e., focusing on text preprocessing and leveraging state-of-the-art NLP algorithms rather than creating ensembles to improve performances. Our extensive evaluation with a large scale SE dataset has identified a model that has 95.8\% accuracy and boosts 88.9\% $F1_1$ score in identifying toxic texts. This model's performances are within 3\% of the best one from a non-SE domain.   This result also suggests that with a carefully labeled large scale dataset, we can train an SE domain specific toxicity detector that achieves performances that are close to those of toxicity detectors from non-SE domains.

\vspace{4pt} \noindent  \textbf{Lesson 2: Performance from Random Forest's optimum configuration may be adequate if GPU is not available.}

While a deep learning-based model (i.e., BERT) achieved the best performances during our evaluations, that model is computationally expensive. Even with a high end GPU such as Titan RTX, our BERT model required on average 1,614 seconds for training.  We found that  \code{RandomForest} based models trained on a Core-i7 CPU took only 64 seconds on average. 

\revised{During a classification task, RF generates the decision using majority voting from all sub trees.  RF is suitable for high dimensional noisy data like the ones found in text classification tasks~\cite{islam2019semantics}. With carefully selected preprocessing steps to better understand contexts (e.g., profanity count) RF may perform well for binary toxicity classification tasks. In our model, after adding profane count features, RF achieved an average accuracy of 95.5\% and $F1_1$- score of 87.9\%, which are within 1\% of those achieved by BERT. Therefore, if computation cost is an issue, a \code{RandomForest} based model may be adequate for many practical applications. However, as our RF model uses a context-free vectorizer, it may perform poorly on texts, where prediction depends on surrounding contexts. Therefore, for  a practical application, a user must take that limitation into account. }

\vspace{4pt}  \noindent \textbf{Lesson 3: Preprocessing steps do improve performances.}

We have implemented five mandatory and three optional preprocessing steps in ToxiCR. The mandatory preprocessing steps do improve performances of our models. For example,  a DPCNN model without these preprocessing achieved 91\% accuracy and 78\% $F1_1$ (Table~\ref{tab:basline}). On the other hand, a model based on the same algorithm  achieved 94.4\% accuracy and 84.5\% $F1_1$ with these preprocessing steps.  Therefore, we recommend using both domain specific and general preprocessing steps. 

\revised{Two of our pre-processing steps are SE domain specific (i.e., Identifier Splitting, Programming Keywords removal).  
Our empirical evaluation of those steps (Section~\ref{sec:preprocess-eval}) suggest that eight out of the ten models (i.e., except SVM and DPCNN) achieved significant performance improvements through these steps. Although, none of the models showed significant degradation through these steps, significant gains were dependent on algorithm selection, with CLE algorithms gaining only from keyword removal and identifier splitting improving only the DNN ones.}

\vspace{4pt} \noindent  \textbf{Lesson 4:  Performance boosts from the optional preprocessing steps are algorithm dependent.}

The three optional preprocessing steps also improved performances of the classifiers. However, performance gains through the these steps were algorithm dependent. The \code{profane-count} preprocessing had the highest influence as nine out of the ten models gained performance with this step. On the other \code{id-split} was the least useful one with only three DNN models gaining minor gains with this step.   
CLE algorithms gained the most with $\approx$ 1\% boost in terms of accuracies and  1 -3\% in terms of $F1_1$ scores. On the other hand, DNN algorithms had relatively minor gains  (i.e., less than 1\%) in both accuracies and $F1_1$ scores.
Since DNN models utilize embedding vectors to identify semantic representation of texts, those are less dependent on these optional preprocessing steps.

\vspace{4pt} \noindent \textbf{Lesson 5:  Accurate identification of self-deprecating texts remains a challenge.}

\revised{Almost 11\% (out of 856 misclassified texts) of the errors from our best performing model were due to self-deprecating texts.}
Challenges in identifying such texts have been also acknowledged by prior toxicity detectors~\cite{Detoxify,wang2020nova,zhang2018conversations}. Due to the abundance of self-deprecating texts among code review interactions~\cite{sarker2020apsec,miller2022toxic}, we believe  that this can be an area to improve on for future SE domain specific toxicity detectors. 

\vspace{4pt} \noindent \revised{ \textbf{Lesson 6: Achieving even higher performance is feasible.} Since 85\% of errors are due to failures of our models to accurately comprehend the contexts of words, we believe achieving further improved performance is feasible. Since supervised models learn better from larger training datasets, a larger dataset (e.g., Jigsaw dataset includes 160K samples), may enable even higher performances. Second,  NLP is a rapidly progressing area with state-of-the-art techniques changing almost every year. Although, we have not evaluated the most recent generation of models, such as GPT-3 \cite{brown2020language} and XLNet \cite{yang2019xlnet} in this study, those may help achieving better performances, as they are better at identifying contexts. 
}

%% file: Sections/threats.tex
\section{Threats to Validity} \label{threats}
In the following, we discuss the four common types of threats to validity for this study.

\subsection{Internal validity}
The first threat to validity for this study is our selection of data sources which come from four FOSS projects. While these projects represent four different domains, many domains
are not represented in our dataset. Moreover, our projects represent some of the top FOSS projects with organized governance. Therefore, several categories of highly offensive texts may be underrepresented in our datasets.

The notion of toxicity also depends on multitude of different factors such as culture, ethnicity, country of origin, language, and relationship between the participants. We did not account for any such factors during our dataset labeling.

\subsection{Construct validity}
Our stratified sampling strategy was based on toxicity scores obtained from the perspective API. Although, we manually verified all the texts classified as ‘toxic’ by the PPA, we randomly selected only (5,510\footnote{Code review 1 dataset} + 9,000 =14,510) texts that had PPA scores of less than 0.5.  Among those 14,510 texts, we identified only 638 toxic ones (4.4\%). If both the  PPA and our random selections missed some categories of toxic comments, instances of such texts may be missing in our datasets. Since our dataset is relatively large (i.e., 19,651 ), we believe that this threat is negligible.

According to our definition, Toxicity is a large umbrella that includes various anti-social behaviour such as offensive names, profanity, insults, threats, personal attacks, flirtations, and sexual references. Though our rubric is based on the Conversational AI team, we have modified it to fit a diverse and multicultural professional workplace such as an OSS project. As the notion of toxicity is a context dependent complex phenomena, our definition may not fit many organizations, especially the homogeneous ones.

Researcher bias during our manual labeling process could also cause mislabeled instances. To eliminate such biases, we focused on developing a rubric first. With the agreed upon rubric, two of the authors  independently labeled each  text and achieved `almost perfect' ($\kappa =0.92$) inter-rater agreement. Therefore, we do not anticipate any significant threat arising from our manual labeling.

We did not change most of the hyperparameters for the CLE algorithms and accepted the default parameters. Therefore, some of the CLE models may have achieved better performances on our datasets through parameter tuning. To address this threat, we used the \code{GridSearchCV} function from the scikit-learn library with the top two CLE models (i.e., \code{RandomForest} and \code{DecisionTree}) to identify the best parameter combinations. Our implementation explored six parameters with total 5,040 combinations for \code{RandomForest} and five parameters with 360 combinations for {DecisionTree}. Our results suggest that most of the default values are identical to those from the best performing combinations identified through \code{GridSearchCV}. We also reevaluated RF and DT with the \code{GridSearchCV} suggested values, but did not find any statistically significant (paired sample t-tests, $p>0.05$) improvements over our already trained models.

For the DNN algorithms, we did not conduct extensive hyperparameter search due to computational costs. However, parameter values were selected based on the best practice reported in the deep learning literature. Moreover, to identify the best DNN models we used validation sets and used \code{EarlyStopping}. Still we may not have been  able to achieve the best possible performances from the DNN models during our evaluations.

\subsection{External validity}
Although, we have not used any project or code review specific pre-processing, our dataset may not adequately represent texts from other projects or other software development interactions such as issue discussions, commit messages, or question /answers on StackExchange. Therefore, our pretrained models may have degraded performances on other contexts. However, our models can be easily retrained using a different labeled datasets from other projects or other types of interactions. To facilitate such retraining, we have made both the source code and instructions to retrain the models publicly available~\cite{replication}.

\subsection{Conclusion validity}
To evaluate the performances our models, we have standard metrics such as accuracy, precision, recall, and F-scores. For the algorithm implementations, we have extensively used state-of-the-art libraries such as scikit-learn~\cite{scikit-learn} and TensorFlow~\cite{tensorflow}.  We also used 10-fold cross-validations to evaluate the performances of each model. Therefore, we do not anticipate any threats to validity arising from the set of metrics, supporting library selection, and evaluation of the algorithms.

%% file: Sections/conclusion.tex
\section{Conclusion and Future Directions} \label{conclusion}
This paper presents design and evaluation of ToxiCR,  a supervised learning-based classifier to identify toxic code review comments. ToxiCR includes a choice to select one of the ten supervised learning algorithms, an option to select text vectorization techniques, five mandatory and three optional processing steps, and a large-scale labeled dataset of 19,651 code review comments. With our rigorous evaluation of the models with various combinations of preprocessing steps and vectorization techniques, we have identified the best combination that boosts 95.8\% accuracy and 88.9\% $F1_1$ score. We have released our dataset, pretrained models, and source code publicly available on Github~\cite{replication}. We anticipate this tool being helpful in combating toxicity among FOSS communities. As a future direction, we aim to conduct empirical studies to investigate how toxic interactions impact code review processes and their outcomes among various FOSS projects.